\documentclass[aps,prr,twocolumn,superscriptaddress,footinbib]{revtex4-2}
\usepackage{amsmath,amssymb,bm}
\usepackage{graphicx}
\usepackage{epstopdf}
\usepackage{latexsym}
\usepackage[normalem]{ulem} 
\usepackage[usenames,dvipsnames]{color}
\usepackage{hyperref}
\usepackage{natbib}
\usepackage{braket}
\usepackage{comment}
\begin{document}

\newcommand{\Rev}[1]{{\color{blue}{#1}\normalcolor}} 
\newcommand{\Com}[1]{{\color{red}{#1}\normalcolor}}

\title{Identifying and harnessing dynamical phase transitions for quantum-enhanced sensing}
\date{\today}

\author{Q. Guan}
\affiliation{Homer L. Dodge Department of Physics and Astronomy, The University of Oklahoma, Norman, Oklahoma 73019, USA}
\affiliation{Center for Quantum Research and Technology, The University of Oklahoma, Norman, Oklahoma 73019, USA}
\author{R.~J. Lewis-Swan}
\affiliation{Homer L. Dodge Department of Physics and Astronomy, The University of Oklahoma, Norman, Oklahoma 73019, USA}
\affiliation{Center for Quantum Research and Technology, The University of Oklahoma, Norman, Oklahoma 73019, USA}

\begin{abstract}
We use the quantum Fisher information (QFI) to diagnose a dynamical phase transition (DPT) in a closed quantum system, which is usually defined in terms of non-analytic behaviour of a time-averaged order parameter. Employing the Lipkin-Meshkov-Glick model as an illustrative example, we find that the DPT correlates with a peak in the QFI that can be explained by a generic connection to an underlying excited-state quantum phase transition that also enables us to also relate the scaling of the QFI with the behaviour of the order parameter. Motivated by the QFI as a quantifier of metrologically useful correlations and entanglement, we also present a robust interferometric protocol that can enable DPTs as a platform for quantum-enhanced sensing.
\end{abstract}

\maketitle  

\section{Introduction} 
The isolation and control of quantum systems at the single-particle level in atomic, molecular and optical platforms has driven a surge of experimental interest in studying non-equilibrium phenomena. As a consequence, it has become clear that non-equilibrium quantum systems that feature coherence, entanglement and correlations, can be important platforms for next-generation quantum technologies \cite{Pezze_2018, RevModPhys.89.035002}. 

From the fundamental perspective, dynamical phase transitions (DPTs) \cite{Heyl_DPTtheory_2013,Schiro10,Sciolla_2011,Halimeh_2017_prethermal,Roos_DPT_2017,Monroe_DPT_2017,Silva_DQPT_2018,Thywissen_DPT_2018,Duan_DPT_2019,Duan_DPT_2020,Muniz2020,Xu_2020} are being pursued in an effort to develop a framework to understand and classify non-equilibrium quantum matter. Here, we focus on DPTs in a closed system, defined as a critical point separating distinct dynamical behaviours (phases) that emerge after a quench of system parameters \cite{lerose2018dpt,schiro2011,Peronaci2015dpt,sciolla_2013,Chiocchetta_2015,Chiocchetta_2017,eckstein09,gambassi,Knap,Lang_2018}, sometimes referred to as DPT-I. Analogous to equilibrium phase transitions, DPTs are characterized by a \emph{time-averaged} order parameter that distinguishes dynamical phases and features non-analytic behaviour at the critical point. A distinct formalism of dynamical \emph{quantum} phase transitions (DQPTs or DPT-II) also exists \cite{Heyl_DPTtheory_2013,Martin2014PRL,Roos_DPT_2017,Silva_DQPT_2018,Oscar2018PRB,Jad2018PRL}, but we do not consider those here. An important current question is to understand the role of entanglement and coherence in DPTs \cite{Monroe_DPT_2017,RLS_Dicke_2021,Sun_2018,Huang_2016} and how these might be harnessed for quantum science applications.

In this manuscript, we theoretically demonstrate that the quantum Fisher information (QFI) \cite{Braunstein1994}, which quantifies metrologically useful entanglement and correlations in a quantum state \cite{Smerzi_2012,Toth2012PRA,Brun_2014}, can be used to characterize DPTs via an underlying connection to excited-state quantum phase transition (EQPTs) \cite{Duan_DPT_2020,RLS_Dicke_2021}. Our method shares analogies with studies of the fidelity susceptibility \cite{You_2007,Gu_2010,Capponi_2010} for ground-state quantum phase transitions (QPTs), but is distinguished by the addition of time as a relevant variable. We employ a numerical study of the dynamical phase diagram of the paradigmatic Lipkin-Meshkov-Glick (LMG) model to illustrate our arguments, and use a semi-analytic model to establish a quantitative connection between the scaling of the QFI and the order parameter. Furthermore, we demonstrate that these quantum signatures of the DPT can also be accessed through a related many-body echo combined with simple global measurements. This latter result, in particular, opens a realistic path for the harnessing of DPTs for quantum-enhanced sensing \cite{Tsang2013}.

\section{Signatures of DPTS in the QFI}
To outline our arguments most generally, consider a many-body Hamiltonian describing a closed system,
\begin{equation}
\label{H_general}
\hat{H}(\lambda)= \hat{H}_0 + \lambda \hat{H}_{1},
\end{equation}
where $[\hat{H}_0,\hat{H}_1] \neq 0$ and $\lambda$ is a tunable parameter. The evolution of an initial state $\vert \psi_0 \rangle$ under $\hat{H}(\lambda)$ is given by $\vert \psi(\lambda, t) \rangle = e^{-i\hat{H}(\lambda) t}\vert \psi_0 \rangle$, and a time-averaged order parameter $\bar{\mathcal{O}} = \frac{1}{T}\int_0^T\langle\hat{\mathcal{O}}(t)\rangle ~ dt$ distinguishes ordered ($\bar{\mathcal{O}} \neq 0$) and disordered ($\bar{\mathcal{O}} = 0$) dynamical phases. A DPT is signaled by non-analytic behaviour in $\bar{\mathcal{O}}$ at a critical point $\lambda_{\mathrm{cr}}$ separating the phases \cite{eckstein09,gambassi,Knap,Lang_2018}.

We propose to characterize the DPT by probing how the state $\vert \psi(\lambda, t) \rangle$ abruptly changes as the system is quenched through $\lambda_{\mathrm{cr}}$. This mirrors uses of the fidelity susceptibility to quantify an abrupt change in the ground-state wavefunction at an equilibrium transition \cite{Quan_2006,Zanardi_2006,You_2007,Gu_2010,Capponi_2010,Troyer_2015}. We similarly define the QFI as the susceptibility \cite{Braunstein1994,Brun_2014,Mirkhalaf_2021},
\begin{equation}
\label{quantum_fisher_info}
F_{Q}(\lambda, t) = -4\frac{\partial^2\mathcal{F}(\lambda,\delta\lambda, t)}{\partial(\delta\lambda)^2} \bigg\vert_{\delta\lambda \to 0}, 
\end{equation}
where $\mathcal{F}(\lambda,\delta\lambda, t) = \vert \langle \psi(\lambda, t) \vert \psi(\lambda+ \delta\lambda, t) \rangle \vert$ is the overlap between two dynamical states that differ by a perturbation $\delta\lambda$ to the driving parameter, equivalent to a Loschmidt echo (LE) \cite{LoschmidtEcho_2016,Prosen2003,GORIN200633} $\mathcal{F}(\lambda,\delta\lambda, t) \equiv \vert \langle \psi_0 \vert e^{i \hat{H}(\lambda) t} e^{-i\hat{H}(\lambda+\delta\lambda)t} \vert \psi_0 \rangle \vert$. About the critical point, $\lambda \approx \lambda_{\mathrm{cr}}$, we expect $\mathcal{F}$ to abruptly decrease as the dynamical states lie in distinct phases and become rapidly orthogonal, and we predict that the DPT is signaled by a corresponding sharp peak in the QFI. This signature complements the time-averaged order parameter and, similar to an equilibrium fidelity susceptibility, has the capacity to carry more information about the system as it based on a state overlap \cite{RLS_MQC_2020}.
Moreover, the QFI could potentially provide a robust, agnostic probe of more sophisticated DPTs without requiring any a priori knowledge of, e.g., the appropriate order parameter for the transition.

This use of the LE for DPTs is distinct from DQPTs, wherein the LE arises as a return fidelity, $\langle \psi_0 \vert e^{-i \hat{H}(\lambda) t} \vert \psi_0 \rangle$, that relates to non-analyticities at a \emph{critical time} rather than parameter. Moreover, our LE compares states related by a small perturbation to the Hamiltonian, whereas the survival fidelity is defined relative to a stationary state $\vert \psi_0 \rangle$ and can be highly non-perturbative.

\section{LMG Model}
We demonstrate the validity of our prediction using the LMG model \cite{LMG_1965, ULYANOV1992179} as a paradigmatic example of a DPT \cite{lerose2019dpt}. Our choice is motivated by the collective nature of the model, which describes an ensemble of $N$ mutually interacting spin-$1/2$ particles subject to transverse and longitudinal fields, as this facilitates a tractable analysis of the dynamics across a range of parameter regimes and system sizes. Moreover, the LMG model has been studied in the context of trapped ions \cite{Monroe_DPT_2017,Roos_DPT_2017}, cavity-QED \cite{Muniz2020} and cold atoms \cite{Zibold2010,PhysRevLett.95.010402,Labuhn_2016,Oberthaler_2015}. It is defined by the Hamiltonian \cite{Muniz2020}
\begin{equation}
\label{LMG}
\hat{H}_{\mathrm{LMG}} =  -\frac{\chi}{N}\hat{S}^2_z - \Omega\hat{S}_x -\omega \hat{S}_z,
\end{equation}
where $\hat{S}_{\alpha} = \sum_{i=1}^N\hat{\sigma}_{\alpha, i}/2$ with $\alpha=x, y, z$ are collective spin operators and $\sigma_{\alpha,i}$ are Pauli matrices for the $i$th spin-$1/2$ particle. The Hamiltonian conserves the total spin, $\hat{S}^2 = \hat{S}_x^2 + \hat{S}_y^2 + \hat{S}_z^2$, and we focus on the maximally collective sector, i.e., states with $\langle\hat{S}^2\rangle = N(N/2+1)/2$.

\begin{figure}[!]
    \centering
    \includegraphics[width=8cm]{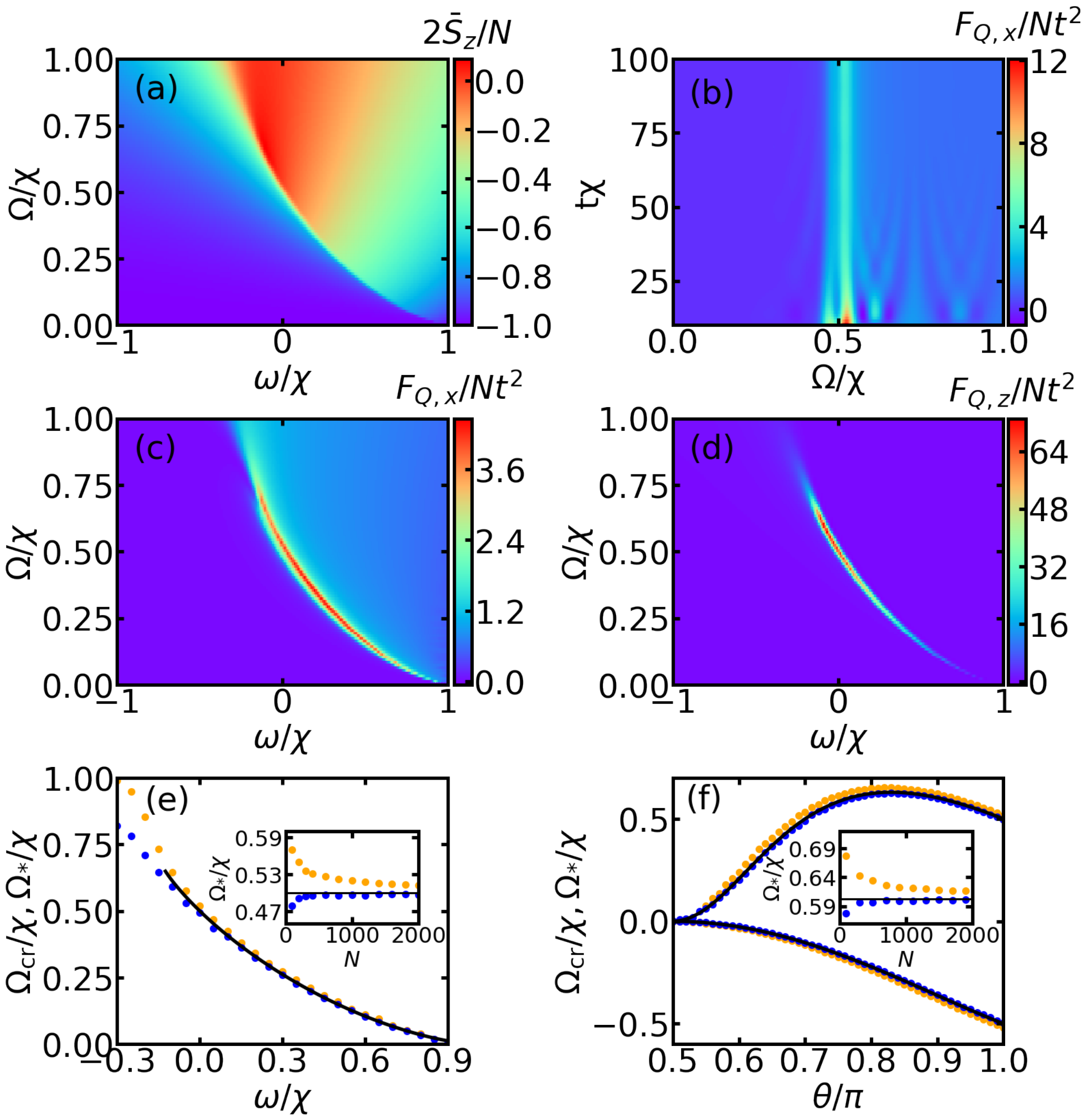}
    \caption{(a) Classical dynamical phase diagram using the time-averaged order parameter $\bar{S}_z$. In the ordered phase $\bar{S}_z \neq 0$ while in the disordered phase $\bar{S}_z = 0$. The initial state is all spins polarized along $-z$. (b) Time evolution of QFI $F_{Q,x}$ as a function of $\Omega/\chi$ and fixed $\omega/\chi = 0$ [initial state as per (a) with $N=10^3$]. (c)-(d) Dynamical phase diagram computed with $F_{Q,x}$ and $F_{Q,z}$ at fixed $\chi t = 10^3$ [other parameters as per (b)]. (e) Phase boundary $\Omega_{\mathrm{cr}}(\omega)$ computed with $\bar{S}_z$ (solid line) compared to $\Omega_{*}(\omega)$ determined from peak values of $F_{Q,x}$ (orange dots) and $F_{Q,z}$ (blue dots) in (c)-(d). (f) Same as (e) but for fixed $\omega/\chi = 0$ and varying the tipping angle $\theta$ (relative to $+z$) of the initial spin state with fixed azimuthal angle $\phi=0.2\pi$. These initial states break the symmetry $\Omega \to -\Omega$ of the phase diagram and so we plot critical points for both positive and negative transverse field strength. Insets of (e) and (f) show scaling of $\Omega_*$ with $N$ for $\omega = 0$ and $\theta = 0.9\pi$, respectively. 
    }
    \label{fig:fig1}
\end{figure}

The dynamical phase diagram in the classical limit ($N\to\infty$) is shown in Fig.~\ref{fig:fig1}(a), for an initial state of all spins polarized along $-\hat{z}$. A pair of dynamical phases are defined in terms of a time-averaged order parameter $\bar{S}_z$ and most easily described in the limit $\omega = 0$: For $\Omega \ll \chi$ interactions force the spins to remain closely aligned to $-\hat{z}$ and $\bar{S}_z \neq 0$, while for $\Omega \gg \chi$ the dynamics is dominated by single-particle Rabi flopping of each spin-$1/2$ about the $+\hat{x}$-axis and thus $\bar{S}_z = 0$. A critical point separates the phases at $\Omega_{\mathrm{cr}}/\chi = 1/2$. Similar analysis holds for $\omega \neq 0$, although the DPT smooths out to a crossover for $\omega/\chi \leq -1/8$ \cite{SM,Muniz2020}. The dynamical phase diagram is symmetric for $\Omega \to -\Omega$ in this case.

In Figs.~\ref{fig:fig1}(b)-(f) we use an efficient Chebyshev expansion algorithm to integrate the dynamics of a system with $N = 10^3$ \cite{Kosloff_JCP1984,SM} and investigate the DPT using the QFI. We consider independent perturbations of the longitudinal $\delta\omega\hat{S}_z$ [$F_{Q,z}(\Omega, \omega, t)$] and transverse $\delta\Omega\hat{S}_x$ [$F_{Q,x}(\Omega, \omega, t)$] fields, and use an equivalent initial state $\vert \psi_0 \rangle = \vert (-N/2)_z \rangle$ where $\hat{S}_z \vert m_z \rangle = m_z \vert m_z \rangle$. Panel (b) shows typical dynamical behaviour of $F_{Q,x}/Nt^2$ as $\Omega/\chi$ is varied and $\omega/\chi = 0$. The normalization is chosen to absorb the expected long-time growth of $F_{Q,x} \propto t^2$. Around the critical point $\Omega_{\mathrm{cr}}/\chi = 1/2$ we observe a pronounced increase in the QFI in both transient and long-time ($\chi t \gg 1$) regimes, such that $F_{Q,x}/Nt^2 \gg 1$. 
Within the transient time regime, the $F_Q/Nt^2$ shows multiple peaks for several medium values of $\chi t$ that are determined numerically~\cite{SM}.
The QFI also distinguishes the ordered phase, $F_{Q,x}/Nt^2 \to 0$, from the disordered phase, $F_{Q,x}/Nt^2 \approx 1$.

Panels (c) and (d) show our key result: The DPT as a function of $\Omega$ and $\omega$ is identified by demonstrable peaks in the QFI at long times $\chi t \gg 1$. The change of the transition to a crossover for $\omega/\chi < -1/8$ is also indicated clearly by a broadened and diminishing peak of the QFI. Panel (e) shows that the critical value $\Omega_{*}$, determined from the peak positions of $F_{Q, x}$ or $F_{Q, z}$, agrees excellently with $\Omega_{\mathrm{cr}}$ determined from $\bar{S}_z$ in the $N\to\infty$ limit, up to finite-size effects (see inset).

Similarly, panel (f) demonstrates that the QFI reproduces the signature dependence of the DPT on the initial state \cite{Muniz2020}. We compare $\Omega_{*}$ and $\Omega_{\mathrm{cr}}$ as a function of the initial state $\vert \psi_0 \rangle = \vert \theta, \phi \rangle$ where $\theta$ is the tipping angle of the collective spin with respect to $\hat{z}$ on the collective Bloch sphere and we fix the azimuthal angle $\phi=0.2\pi$. For these initial states the phase diagram is no longer symmetric for $\Omega \to -\Omega$, so we plot both relevant values of the critical transverse field (fixing $\omega/\chi = 0$).

\section{Critical scaling of the QFI} The scaling of the QFI with system size is important for identifying how any generated correlations and entanglement can be useful for quantum sensing. Specifically, the QFI provides a lower bound for the accuracy to which the driving parameter $\delta\lambda$ can be determined, $\Delta\lambda\geq 1/\sqrt{F_{Q}(\lambda, t)}$ \cite{Braunstein1994}. The standard quantum limit, e.g., the sensitivity that can be attained with quasi-classical uncorrelated states, sets a bound $(\Delta \lambda)^2_{\mathrm{SQL}} \geq  1/(Nt^2)$ or equivalently $F^{\mathrm{cl}}_{Q,(x,z)} \leq Nt^2$ \cite{Giovannetti_2006,Skotiniotis_2015,SM}. Supralinear scaling of the QFI with $N$ near the critical point would indicate potential uses for DPTs in quantum-enhanced sensing. 

\begin{figure}[!]
    \centering
    \includegraphics[width=8cm]{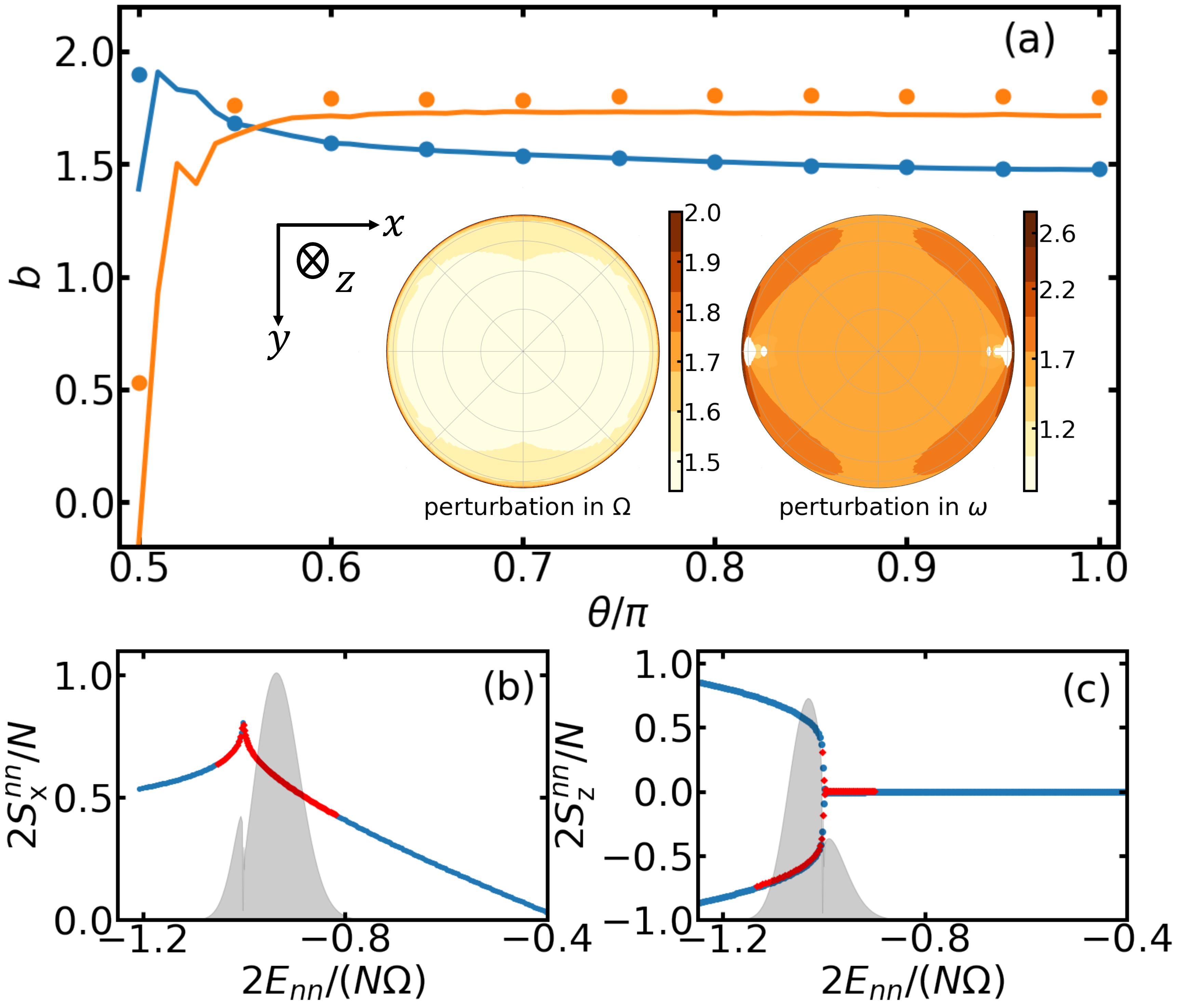}
    \caption{(a) Scaling exponent of $F_{Q}(\Omega_*)$ at fixed $\omega/\chi = 10^{-4}$. 
    We fit $F_Q(\Omega_*) \sim aN^b$ across a window of $N \in [100, 2000]$ for different initial states parameterized by $\theta$ and $\phi = 0$. Data from numerical integration of the dynamics to $\chi t = 10^3$ (lines) is compared to the analytic expression $F_Q^{\mathrm{sec}}$ [Eq.~(\ref{long_time_F})] for transverse (blue data) and longitudinal (orange data) perturbations. Inset: Scaling exponent for complete range of initial polarized states in the southern hemisphere (results are symmetric about the equator) obtained from $F_Q^{\mathrm{sec}}$. (b)-(c) $S_{x,z}^{nn}$ as a function of eigenenergy $E_{nn}$, obtained by numerical diagonalization of $\hat{H}$ at $\Omega = \Omega^*$ and $\omega = 10^{-4}$ for $N=10^3$. Shaded background indicates the distribution $\vert c_n \vert^2$ of the initial state in the eigenbasis [also indicated as red highlights on $S_{x,z}^{nn}$ to show relevant contributions of $S_{x,z}^{nn}$ in Eq.~(\ref{long_time_F})]. 
    }
    \label{fig:fig2}
\end{figure}

We empirically fit the maximum of $F_Q$ over $\Omega$ for fixed $\omega/\chi = 10^{-4}$ and $\chi t \gg 1$, i.e., $F_Q(\Omega^*) = aN^b$. A finite $\omega$ is chosen to purposely break a parity symmetry of the Hamiltonian \cite{Ribeiro_2008,SM}. Figure \ref{fig:fig2}(a) shows the exponent $b$ as a function of the tipping angle $\theta$ of the initial state with fixed $\phi=0$. Excluding a region near the equator ($\theta \approx \pi/2$) we observe approximately constant values of $b \approx 1.5$ and $1.75$ for $F_{Q,x}$ and $F_{Q,z}$, respectively, indicating the presence of metrologically useful correlations and entanglement for sub-SQL sensing, e.g., $(\Delta \lambda)^2 =  F_Q < (\Delta \lambda)^2_{\mathrm{SQL}}$ for large $N$, and which cannot be captured by mean-field theory~\cite{SM}.

To understand the scaling of the QFI we consider the generic Hamiltonian $\hat{H} = \hat{H}_0 + \lambda\hat{H}_1$ and use an approximate analytic expression for the long-time secular contribution \cite{Brun_2014,SM},
\begin{align}
\label{long_time_F}
F^{\mathrm{sec}}_{Q}(\lambda, t) \approx 
4t^2\left[ \sum_{n}  \vert c_n\vert^2 \left( H_{1}^{nn} \right)^2  - \left(\sum_{n}  \vert c_n \vert^2 H_{1}^{nn} \right)^2 \right].
\end{align}
Here, $\vert n \rangle$ are the eigenstates of $\hat{H}$, $c_n = \langle n \vert \psi_0  \rangle$ is the projection of the initial state into the eigenbasis and $H_{1}^{nn}=\langle n|\hat{H}_{1}|n\rangle$. Equation~\eqref{long_time_F} is valid when the spectrum of $\hat{H}$ is non-degenerate or in the case where the degenerate eigenstates occupy different sectors of a symmetry which leaves $\hat{H}_1$ invariant \cite{SM}. Numerical evaluation of Eq.~(\ref{long_time_F}) for the LMG model ($\hat{H}_1 = \hat{S}_z$ or $\hat{S}_x$) agrees well with our numerical simulations of the dynamics for $\theta$ away from the equatorial plane. Note we chose a small $\omega/\chi = 10^{-4}$ to ensure the spectrum is non-degenerate for the case of $F^{\mathrm{sec}}_{Q,z}$, although this is not crucial to our results \cite{SM}. Minor disagreement between $F^{\mathrm{sec}}_{Q,z}$ and $F_{Q,z}$, and similarly for $F^{\mathrm{sec}}_{Q,x}$ and $F_{Q,x}$ near the equator, is due to corrections from transient terms ignored in Eq.~(\ref{long_time_F}). 

Equation~(\ref{long_time_F}) gives a simple interpretation for the long-time QFI: A peak in the QFI is a result of enhanced fluctuations in $H_{1}^{nn}$ attributable to either a sharp change in the properties of the eigenstates or the projection of the initial state into the eigenbasis, both of which can be correlated with the emergence of a DPT \cite{Dag_2018,Puebla_2014}. For many systems~\cite{RLS_Dicke_2021,Puebla2020,Duan_DPT_2020}, including the LMG model~\cite{Puebla_2014}, the DPT is triggered by the former effect due to an EQPT which leads to non-analytic features in $S^{nn}_x$ and $S^{nn}_z$ \cite{Ribeiro_2008,Santos_2016}, as shown in Fig.~\ref{fig:fig2}(b) and (c) for a representative calculation with $N=1000$. A sharp cusp (kink) is observed in $S^{nn}_x$ ($S^{nn}_z$) at a critical energy $E_{\mathrm{cr}}/N = -\Omega/2$ for $\Omega/\chi < 1$. This is precisely the average energy of the initial state, $E_0 = \langle \psi_0 \vert \hat{H} \vert \psi_0 \rangle$, as $\Omega/\chi$ (or also $\omega/\chi$ in general) is tuned through the DPT at $\Omega_{\mathrm{cr}}/\chi \approx 1/2$. Thus, the non-analytic behaviour of the DPT already corresponds closely with that of $S_z^{nn}$ via the relation $\bar{S}_z \equiv \sum_n \vert c_n \vert^2 S_z^{nn}$ at long times \cite{Puebla_2014}. 

By examining the projection of the initial state ($c_n$) for $\Omega \approx \Omega_{\mathrm{cr}}$ in Figs.~\ref{fig:fig2}(b) and (c), we conclude that the distribution of relevant $S^{nn}_x$ ($S^{nn}_z$) in Eq.~\eqref{long_time_F} straddles the cusp (kink) and leads to the sudden increase of the QFI at the DPT. Moreover, Eq.~\eqref{long_time_F} combined with the approximations: i) $S^{nn}_{x,z}/N \approx A_{x,z}+B_{x,z}(E/N -E_{\mathrm{cr}}/N)^{\gamma_{x,z}}$ near $E\approx E_{\mathrm{cr}}$ \cite{Perez_2011}, and ii) the energy fluctuations of an initial coherent spin state typically scale as $\Delta E \sim \sqrt{N}$, allows us to qualitatively predict the scaling of the QFI as $F_{Q}/t^2 \sim N^{2-\gamma_{x,z}}$ \cite{SM}. From numerical diagonalization of the LMG Hamiltonian we obtain $\gamma_x \simeq 0.495\pm0.005$ and $\gamma_z \simeq 0.25\pm0.02$ which is consistent with the approximate scaling of $F_{Q,x}/t^2 \sim N^{1.5}$ and $F_{Q,z}/t^2 \sim N^{1.75}$. The scaling of $F_{Q,z}$ is intimately related to the scaling of the order parameter $\bar{S}_z \sim (\Omega-\Omega_{\mathrm{cr}})^{\gamma_z}$ [care of approximation (i)] for a finite size system~\footnote{In the classical ($N\to\infty$) limit the DPT is characterized by a logarithmic divergence, $\bar{S}_z \sim 1/\mathrm{log}(\Omega_{\mathrm{cr}}-\Omega)$. However, for the system sizes we probe the divergence is dominated by finite-size effects and thus a power-law is suitable (see Ref.~\cite{SM}).}. Thus the QFI, which is a detailed measure of how rapidly the dynamical state changes across the DPT, is intuitively governed by the sharpness of the DPT in terms of the time-averaged order parameter. 

Our analysis further supports that the QFI correctly diagnoses the DPT. For $N\to \infty$ the relative energy fluctuations $\Delta E/E$ of the initial state vanish and $F^{\mathrm{sec}}_{Q,z}$ ($F^{\mathrm{sec}}_{Q,x}$) will have large contributions from $S^{nn}_z$ ($S^{nn}_x$) at $E_0 \to E_{\mathrm{cr}}^-$ ($E_0 \to E_{\mathrm{cr}}^+$) [consistent with observations in Fig.~\ref{fig:fig1}(e) that show $\Omega_{*}$ computed from the QFI approaching $\Omega_{\mathrm{cr}}$ from below (above)].

We note that while we understand the connection between the QFI and DPT as being facilitated by the EQPT, it should be distinguished from the latter as being a non-equilibrium result. For example, while EQPTs also feature a divergent fidelity susceptibility (and thus associated QFI), this is a property of eigenstates of an equilibrium system. In contrast, Eq.~\eqref{long_time_F} would give a QFI of zero for an eigenstate of the system, as it is specifically the fluctuations of the initial state (i.e., distribution across the eigenstates) that drives the large QFI and, moreover, are inextricably related to the scaling we observe. Furthermore, directly harnessing the QFI of an EQPT for quantum sensing would be faced with the tandem challenges of controllably preparing an excited state of a many-body system and adiabatically tuning system parameters through a phase boundary. The latter inevitably becomes difficult with increasing system size as the required time scales diverge as a result of the vanishing energy gap.

\section{Practical implementation and application to metrology} 

The QFI can be measured by implementing a LE sequence that also serves as an optimal metrological protocol to characterize $\delta\lambda$ \cite{Sekatski2015,Macri_2016}. Inspecting the formal definition of the QFI, given in Eq.~\eqref{quantum_fisher_info_s}, it is clear that the state overlap $\langle \psi_{\lambda+\delta\lambda} \vert \psi_{\lambda} \rangle = \langle \psi_0 \vert e^{i\hat{H}(\lambda+\delta\lambda)t} e^{-i\hat{H}(\lambda)t} \vert \psi_0 \rangle$ can be obtained by: i) prepare $\vert \psi_0 \rangle$, ii) evolve forward with the unperturbed Hamiltonian $\hat{H}(\lambda)=\hat{H}_0+\lambda \hat{H}_1$ for time $t$, iii) evolve ``backward'' with the perturbed Hamiltonian $-\hat{H}(\lambda+\delta\lambda)=-\hat{H}_0-\lambda \hat{H}_1 -\delta\lambda \hat{H}_1$ for time $t$, and iv) obtain $\mathcal{F}$ by measuring the overlap $\vert \langle \psi_0 \vert \psi_f \rangle \vert^2$ where $\vert \psi_f \rangle \equiv e^{i\hat{H}(\lambda+\delta\lambda)t}e^{-i\hat{H}(\lambda)t}\vert \psi_0 \rangle$. The capability to invert the sign of a Hamiltonian has been demonstrated or proposed in a range of AMO quantum simulators \cite{Linnemann_2016,Martin2017_OTOC,Swingle_2016,Hosten_2016,Li_2017}.

\begin{figure}[!]
    \centering
    \includegraphics[width=8cm]{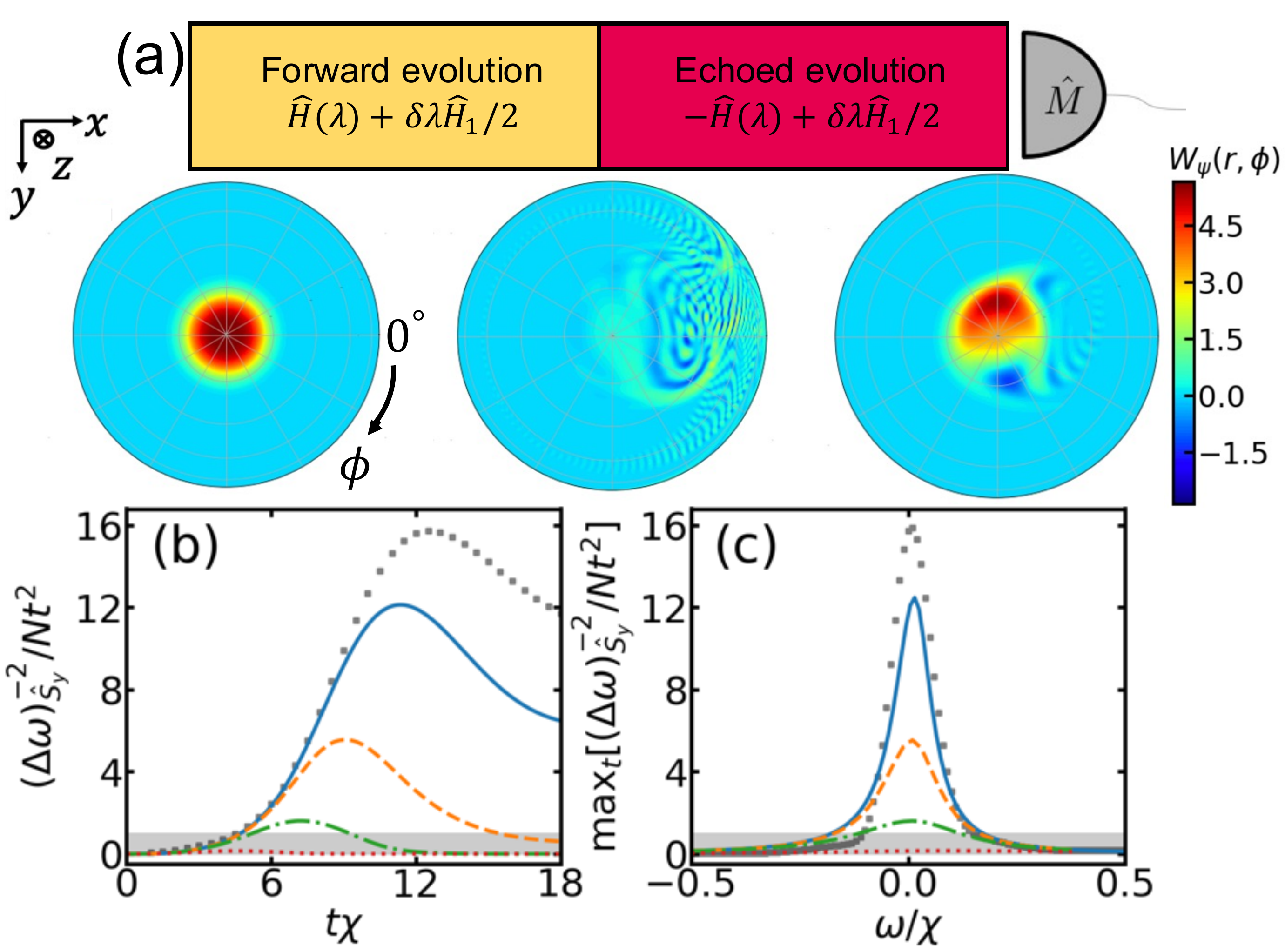}
    \caption{(a) Schematic of echo protocol to obtain QFI/estimate the classical parameter $\lambda$. Typical Wigner functions \cite{WignerDis1994,SM} $W_{\psi}(r, \phi)$ of the initial ($\vert \psi_0\rangle$), intermediate ($\vert \psi(t) \rangle$) and final ($\vert \psi_f \rangle$) states for $\chi t = 12$ and $\delta\omega/\chi = 2\times10^{-3}$ are shown. We plot with polar coordinates $r=\left(1+2S_z/N\right)^{1/4}$ and $\phi = \mathrm{atan}(S_y/S_x)$. (b) Evolution of the inverse sensitivity $(\Delta\omega)^{-2}_{\hat{S}_y}$ for $\Omega/\chi = 1/2$, $\omega/\chi = 0$ and a range of decoherence rates $\Gamma/\chi$. The blue solid, orange dashed, green dot-dashed, and red dotted lines correspond to $\Gamma/\chi = 0, 10^{-3}, 10^{-2}$, and $10^{-1}$, respectively. We also include the QFI $F_{Q,z}$ for the same parameters and $\Gamma/\chi = 0$. (c) Maximum of the normalized inverse sensitivity $\mathrm{max}_t[(\Delta\omega)^{-2}_{\hat{S}_y}/(Nt^2)]$ optimized over time, as a function of $\omega/\chi$ and other parameters as per (b). We also plot the maximum of the normalized QFI, $\mathrm{max}_t[F_{Q,z}/(Nt^2)]$, as an indicator of the DPT. In both (b) and (c) the grey shaded region indicates the regime bounded by the normalized SQL, $(\Delta\omega)^{-2}_{\mathrm{SQL}}/Nt^2 = 1$. All calculations are for $N=100$. 
    }
    \label{fig:fig3}
\end{figure}

We note that the above protocol assumes temporal control over \emph{both} $\lambda$ and $\delta\lambda$. This is a reasonable approach if the goal of the protocol is simply to characterize the QFI and thus the DPT. However, this assumption is not suitable from the perspective of using the DPT for metrology, where the perturbation $\delta\lambda$ is intrinsically unknown. As a result, it is also reasonable for us to focus on the most general scenario where one does not have temporal control of $\delta\lambda$. Thus, we consider an alternative, but entirely equivalent, echo sequence where the perturbation is always present [Fig.~\ref{fig:fig3}(a)]: i) prepare $\vert \psi_0 \rangle$, ii) evolve with $\hat{H}(\lambda+\delta\lambda/2)=\hat{H}(\lambda)+\delta\lambda\hat{H}_1/2$ for time $t$, iii) evolve with $-\hat{H}(\lambda-\delta\lambda/2)=-\hat{H}(\lambda)+\delta\lambda\hat{H}_1/2$ for time $t$, and iv) measure the overlap $\vert \langle \psi_0 \vert \psi_f \rangle \vert^2$, where $\vert \psi_f \rangle \equiv e^{i\hat{H}(\lambda+\delta\lambda/2)t}e^{-i\hat{H}(\lambda-\delta\lambda/2)t}\vert \psi_0 \rangle$, to obtain $\mathcal{F}$. This adapted protocol only requires that the sign of the control parameter $\lambda$ and $\hat{H}_0$ can be varied, while the perturbation $\delta\lambda$ is identically present in both periods of evolution. 

Our protocol is relevant for estimating the small perturbation $\delta\lambda$ when one separately assumes that $\lambda$ is well characterized and independently calibrated to suitably good precision. This is well motivated by the fact that while $\delta\lambda$ and $\lambda$ contribute to the Hamiltonian via the term $\hat{H}_1$, they may be generated by different physical effects. For example, in the case of the LMG model $\hat{H}_1 = \hat{S}_z$ corresponds to an energy splitting in an ensemble of two-level systems, which could be generated via distinct physical mechanisms such that there is some well controlled part $\lambda$ and some unknown perturbation $\delta\lambda$ due to stray external fields. This is no different to an assumption widely used in different metrological scenarios where one tunes a sensor to work at an optimal point [e.g., using a phase offset in an SU(2) or SU(1,1) interferometer]. 

In addition, one could also envision $\delta\lambda$ as not being generated by a separate effect, but instead as some intrinsic uncertainty on top of the control parameter $\lambda$ due to technical noise. In this case, the operation of ``flipping the sign" of $\lambda$ by turning some external control knob would be imperfect and actually realized as $\lambda + \delta\lambda/2 \to -\lambda + \delta/2$~\footnote{An example is Ref.~\cite{gilmore2021quantum}, where this type of error (a fluctuating parameter in the Hamiltonian) explicitly occurs in a trapped ion quantum simulator.}, which follows our proposed echo sequence. Our sensing protocol could then be used to estimate the magnitude of this uncertainty $\delta\lambda$.

While $\mathcal{F}$ is important to compute the QFI, it is also an an optimal signal to infer $\delta\lambda$ \cite{Macri_2016}. Nevertheless, while measuring the final state overlap can be simplified by the fact that DPTs are typically studied with simple uncorrelated initial states, technical challenges, such as the detection resolution required to adequately discriminate states, can still pose a practical hurdle for many platforms.

We can overcome this problem by noting that for a simple (e.g., Gaussian) initial state we expect the final state after the echo to be distinguishable by relatively simple and robust observables such as mean spin projection or occupation \cite{Davis_2016,Hosten_2016,Nolan2017IBR,Haine2018IBR}. Specifically, using the quantum Cramer-Rao bound, measurement of an observable $\hat{M}$ leads to a lower bound for the QFI, $(\Delta\lambda)^{-2}_{\hat{M}} = |\partial_{\delta\lambda}\langle \hat{M}\rangle|^2/\mathrm{var}(\hat{M}) \leq F_Q$, \cite{Braunstein1994,Sekatski2015} which can be made arbitrarily tight for a judiciously chosen observable. For the LMG model, and assuming an initial state with all spins orientated along $-\hat{z}$, the final state $\vert \psi_f \rangle$ after the echo [see the $W_{\psi}(r,\phi)$~\cite{WignerDis1994} in Fig.~\ref{fig:fig3}(a)] approximates a weakly displaced coherent spin state and so is distinguishable from $\vert \psi_0 \rangle$ by measurement of $\hat{M} = \hat{S}_y$ for either choice of perturbation.

In Fig.~\ref{fig:fig3}(b) we compare $(\Delta\omega)^{-2}_{\hat{S}_y}$ to the QFI for a perturbation $\delta\omega$ and a moderate system size $N = 100$ (pertinent for, e.g., trapped ion quantum simulators \cite{Bollinger_2018,Monroe_DPT_2017} and tailored to our later discussion of decoherence). We observe that $\hat{S}_y$ is sufficient to qualitatively replicate the transient features and long-time growth $\propto t^2$ of the QFI. In panel (c) we plot the maximum of $(\Delta\lambda)^{-2}_{\hat{S}_y}/Nt^2$ as a function of $\omega/\chi$ and demonstrate that it qualitatively reproduces the peak in the transient maximum of $F_{Q,z}/Nt^2$ near $\omega \approx 0$, which identifies the DPT. In fact, near the DPT we find $(\Delta\Omega)^{-2}_{\hat{S}_y}/t^2 \sim N^{1.4}$ and $(\Delta\omega)^{-2}_{\hat{S}_y}/t^2 \sim N^{1.78}$ \cite{SM}, closely following the scaling of the QFI. Combined with the observation that $(\Delta\omega)^{2}_{\hat{S}_y} < (\Delta\omega)^{2}_{\mathrm{SQL}} = 1/(Nt^2)$ near the DPT, our results suggest that DPTs could be realistically harnessed for quantum-enhanced sensing by combining dynamical echoes with simple collective measurement observables.  

We also probe the robustness of our results to typical sources of single-particle decoherence. Using the permutation symmetry of the LMG Hamiltonian we are able to efficiently simulate the dynamics of $N=100$ qubits subject to single-particle dephasing at rate $\Gamma$~\cite{Geremia_2010,SM}. For weak decoherence $\Gamma/\chi \lesssim 10^{-2}$ (within reach of, e.g., current state-of-the-art trapped ion quantum simulators \cite{Bollinger_2018}) strong signatures of the DPT remain in $(\Delta\omega)^{-2}_{\hat{S}_y}$, even though we become limited to transient time-scales. Moreover, the sub-SQL sensitivity near the DPT remains robust in the same regimes.

\section{Conclusion}
We have theoretically demonstrated that the QFI can be used to diagnose DPTs. While we establish a semi-quantitative understanding of the QFI via an underlying connection to EQPTs, our analysis demonstrates it is an intrinsically non-equilibrium effect. Despite our focus on the LMG model our results can be widely applicable and it will be interesting to apply our analysis to a broader range of known DPTs \cite{Duan_DPT_2020,Puebla2020,RLS_Dicke_2021,Galitski_2021}. Moreover, our interferometric protocol combining dynamical echoes and measurement of simple observables demonstrates that DPTs could be a promising path for sub-SQL sensing in non-equilibrium many-body systems~\cite{Fiderer2018,Brewer_2019,Nakamura889,Pedrozo2020}, and one which sidesteps typical challenges, such as divergent time-scales, associated with quantum sensors based on equilibrium QPTs.

\begin{acknowledgments}
We acknowledge helpful discussions with D.~Barberena. Numerical calculations for this project were performed at the OU Supercomputing Center for Education \& Research (OSCER) at the University of Oklahoma.
\end{acknowledgments}

\appendix

\section{Quantum Fisher information}
In this section we give several useful expressions for the QFI in the context of general Hamiltonian dynamics. Our discussion includes relevant details connecting the scaling of the QFI to features of the energy spectrum and we also provide an illustrative proof of the upper bound of the QFI for uncorrelated spin states.

\subsection{Exact expression for QFI and secular contributions}
\label{analytical_QFI}
We study the QFI defined as the susceptibility with respect to a small perturbation $\delta\lambda$ of the Hamiltonian $\hat{H} = \hat{H}_0 + \lambda \hat{H}_1$ \cite{Braunstein1994,Brun_2014,Taddei_2013,Mirkhalaf_2021},
\begin{equation}
\label{quantum_fisher_info_s}
F_{Q}(\lambda, t) = -4\frac{\partial^2\mathcal{F}(\lambda,\delta\lambda, t)}{\partial(\delta\lambda)^2} \bigg\vert_{\delta\lambda \to 0} ,
\end{equation}
where $\mathcal{F}(\lambda,\delta\lambda, t) = \vert \langle \psi(\lambda, t) \vert \psi(\lambda+ \delta\lambda, t) \rangle \vert$.
Using the identities 
\begin{equation}
\begin{gathered}
    \mathcal{F}(\lambda, \delta\lambda, t) = \sqrt{\langle\psi(\lambda, t)|\psi(\lambda+\delta\lambda, t)\rangle\langle\psi(\lambda+\delta\lambda, t)|\psi(\lambda, t)\rangle} , \\
    \mathcal{F}(\lambda, \delta\lambda=0, t)=1
\end{gathered}
\end{equation}
and the chain rule
alternatively, we can re-express the QFI as~\cite{Pezze_2018}
\begin{align}
\label{echo}
F_Q (\lambda, t)= 4 \Bigg(\left\langle\frac{d\psi(t,\lambda+\delta\lambda)}{d\delta\lambda}\Big|\frac{d\psi(t,\lambda+\delta\lambda)}{{d\delta\lambda}}\right\rangle \\\nonumber
- \left|\frac{d\langle\psi(t, \lambda)|\psi(t,\lambda+\delta\lambda)\rangle}{d\delta\lambda}\right|^2\Bigg)\Bigg|_{\delta\lambda=0},
\end{align}
which, after invoking the identity \cite{Brun_2014}
\begin{align}
\label{derivative_operator}
&\exp\left[i\hat{H}(\lambda)t\right]\frac{d}{d\lambda}\exp\left[-i\hat{H}(\lambda)t\right] = \\\nonumber
&-i \int_0^tdt'\exp\left[i\hat{H}(\lambda)t'\right]\frac{d\hat{H}}{d\lambda}\exp\left[-i\hat{H}(\lambda)t'\right],
\end{align}
leads to an expression for the QFI in terms of the variance of a time-averaged generator $\overline{(\hat{\mathcal{H}}_1)}_t=\frac{1}{t}\int_0^t \hat{H}_1(t')dt'$ \cite{Brun_2014,Skotiniotis_2015},
\begin{eqnarray}
\label{F_Q_G}
F_Q(\lambda, t) = 4t^2\left[\langle \overline{(\hat{\mathcal{H}}_1)}_t^2 \rangle - |\langle \overline{(\hat{\mathcal{H}}_1)}_t \rangle|^2\right] .
\end{eqnarray}

It is possible to extract a useful expression for the long-time secular behaviour of the QFI  by evaluating the expectations in Eq.~\eqref{F_Q_G} using an expansion of the initial state $\vert \psi_0 \rangle$ of the system over the eigenbasis $\vert n \rangle$ of $\hat{H}$. Specifically, we plug $\vert \psi_0 \rangle = \sum_n c_n \vert n \rangle$ with $c_n \equiv \langle n \vert \psi_0 \rangle$ into Eq.~\eqref{F_Q_G} to find 
\begin{align}
\label{diagonal_0}
& F_Q(\lambda, t) = t^2\Bigg[\sum_{n, k, m}c_n^*c_m\mathrm{e}^{\frac{i\Delta_{nm}t}{2}}H_1^{nk}H_1^{km}\mathrm{sinc}\left(\frac{\Delta_{nk}t}{2}\right)\times \\\nonumber
&\mathrm{sinc}\left(\frac{\Delta_{km}t}{2}\right)-\left|\sum_{n,m}c_n^*c_m H_1^{nm}\mathrm{e}^{\frac{i\Delta_{nm}t}{2}}\mathrm{sinc}\left(\frac{\Delta_{nm}t}{2}\right)\right|^2\Bigg]
\end{align}
where $\Delta_{nm}=E_{nn}-E_{mm}$ for $E_{nn} = \langle n \vert \hat{H} \vert n \rangle$.

In the limit of $t\rightarrow\infty$, 
the $\mathrm{sinc}$ function enforces that only terms with $\Delta_{nk}=\Delta_{km}=0$ survive in Eq.~\eqref{diagonal_0}, leading to 
\begin{align}
\label{diagonal_1}
\lim_{t\rightarrow\infty}F_Q(\lambda, t) = 4t^2\Bigg[\sum_{E_m=E_n=E_k} c_m^*c_n  H_1^{mk}H_1^{kn}\\\nonumber
-\left(\sum_{E_m=E_n}c_m^*c_n H_1^{mn}\right)^2\Bigg].
\end{align}
Assuming the spectrum of $\hat{H}$ is non-degenerate (see discussion below), Eq.~\eqref{diagonal_1} can be reduced to a single sum,
\begin{align}
\label{diagonal_2}
F_Q^{\mathrm{sec}}(\lambda, t) \approx 4t^2\Bigg[\sum_{n} |c_n|^2  |H_1^{nn}|^2
-\left(\sum_{n}|c_n|^2 H_1^{nn}\right)^2\Bigg],
\end{align}
in which we have neglected all the sub-leading order terms due to finite time. As a result, an obvious but important condition for the validity of Eq.~\eqref{diagonal_2} is thus that the coefficient of the $t^2$ term is non-vanishing, such that at sufficiently long times we can justifiably ignore those transient contributions of Eq.~\eqref{diagonal_1}. 

The validity of Eq.~\eqref{diagonal_2} does extend to the case where $\hat{H}$ possesses a degenerate spectrum, if the degenerate states occupy different sectors of a symmetry that leaves $\hat{H}_1$ invariant. Considering the LMG model illustrates this statement clearly: Degenerate pairs of eigenstates occur in the case $\omega=0$ in the self-trapped phase due to the fact that the Hamiltonian possesses a parity symmetry generated by the spin flip operator $\hat{P} = \Pi_{i=1}^N\hat{\sigma}_{x,i}$. We denote the $n$-th pair of degenerate eigenstates by $|+,n\rangle$ and $|-,n\rangle$ where the ``$\pm$" labels the eigenvalues of $\hat{P}$. For $F_{Q,x}$ we have that $[\hat{P},\hat{S}_x]$ commute and so the degenerate terms in Eq.~\eqref{diagonal_1} vanish and only the terms with $\langle \pm, n \vert \hat{S}_x \vert \pm, n \rangle$ survive and lead to Eq.~\eqref{diagonal_2}. On the other hand, for $F_{Q,z}$ one should instead consider eigenstates that mix parity, $(|+,n\rangle\pm|-,n\rangle)/\sqrt{2}$, and for which the symmetric/anti-symmetric combination are uncoupled by $\hat{S}_z$ and thus Eq.~\eqref{diagonal_2} again applies. So that we can consider only a single basis when discussing Eq.~\eqref{diagonal_2}, we include a small $\omega/\chi =10^{-4}$ to break the degeneracy when discussing results for $\omega \to 0$. We have confirmed using numerical calculations that the results obtained with $\omega/\chi =10^{-4}$ for Eq.~\eqref{diagonal_2} match excellently with those of a full numerical computation of $F_{Q,z}$ for arbitrarily small $\omega$, and the special point $\omega = 0$ does not change any of our qualitative conclusions.

\begin{figure}
    \centering
    \includegraphics[scale=0.32]{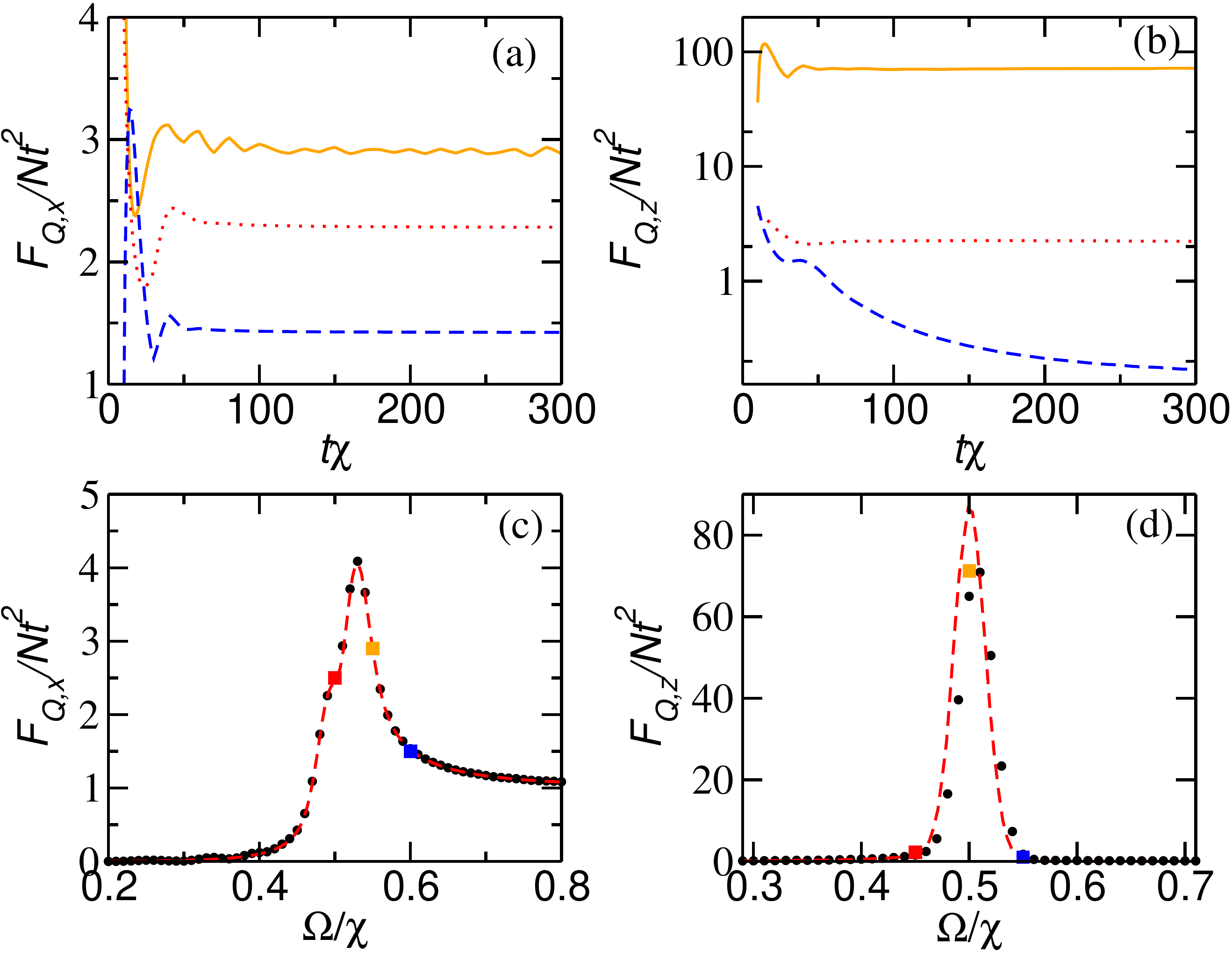}
    \caption{$F_{Q,x/z}/Nt^2$ obtained using numerical time propagation and the diagonal ensemble formula in Eq.~\eqref{diagonal_2}.
    (a/b) The $t^2$ scaling of $F_{Q,x/z}$ for $\omega=10^{-4}$ and $\Omega/\chi=0.5$ (0.45) (red dotted line), $0.55$ (0.5) (orange solid line), and $0.6$ (0.55) (blue dashed line).
    (c)-(d) The $F_{Q, x/z}/Nt^2$ evaluated with $N=1000$, $t\chi=1000$, and $\omega/\chi=10^{-4}$.
    The red dashed lines and black dots in (c) and (d) correspond to the diagonal ensemble results and the time propagation results, respectively. 
    The squares in (c) and (d) mark the three $\Omega$ values shown in (a) and (b) with the same color coding, respectively. 
    }
    \label{fig:sup2}
\end{figure}

Panels (a) and (b) of Fig.~\ref{fig:sup2} show typical time-traces of the normalized QFI $F_{Q,x}/(Nt^2)$ and $F_{Q,z}/(Nt^2)$ for an initial state with all spins polarized along $-\hat{z}$. In the long time limit, $F_{Q,x}/(Nt^2)$ approaches a constant for all parameters, consistent with the form of Eq.~\eqref{diagonal_2}. On the other hand, we find that $F_{Q,z}$ does not demonstrate $t^2$ scaling for certain parameter regimes. For example, in the limit of large $\Omega \gg \chi, \omega$ the Hamiltonian is dominated by the contribution of the transverse field, $\hat{H} \approx -\Omega\hat{S}_x$, and we expect the energy eigenbasis to be close to the eigenstates of $\hat{S}_x$. Hence, to leading order the diagonal matrix elements $S_z^{nn}$ vanish for all $n$. As shown in Fig.~\ref{fig:sup2}(b), we only observe $F_{Q,z} \propto t^2$ when $\Omega \lesssim \Omega_{\mathrm{cr}}$ [red dotted and orange solid lines in Fig.~\ref{fig:sup2}(b)]. For $\Omega\gg\Omega_{\mathrm{cr}}$ [blue dashed line in Fig.~\ref{fig:sup2}(b)], the behaviour of $F_{Q,z}$ is dominated by transient corrections ignored in Eq.~\eqref{diagonal_2} at the time-scales we can probe. 

Consistent with this discussion, Eq.~\eqref{diagonal_2} captures the dynamical phase diagram at long-times in almost perfect agreement with $F_{Q,x}$ (obtained via numerical calculation of the full dynamics). Conversely, while Eq.~\eqref{diagonal_2} does not completely match $F_{Q,z}$ at long times for $\Omega\gtrsim \Omega_{\mathrm{cr}}$ it nevertheless still captures the signatures of the DPT in $F_{Q, z}$. In both cases, we find the scaling of the QFI with system size near the DPT, $\Omega \approx \Omega_{\mathrm{cr}}$, is well captured. Specifically, by directly computing Eq.~\eqref{diagonal_2} in a window of $N \in [100,2000]$ we obtain $F^{\mathrm{sec}}_{Q,x} \sim N^{1.5}$ and $F^{\mathrm{sec}}_{Q,x} \sim N^{1.75}$ which closely agree with results obtained from $F_{Q,x}$ and $F_{Q,z}$ for the same system sizes.

\begin{figure}
    \centering
    \includegraphics[scale=0.32]{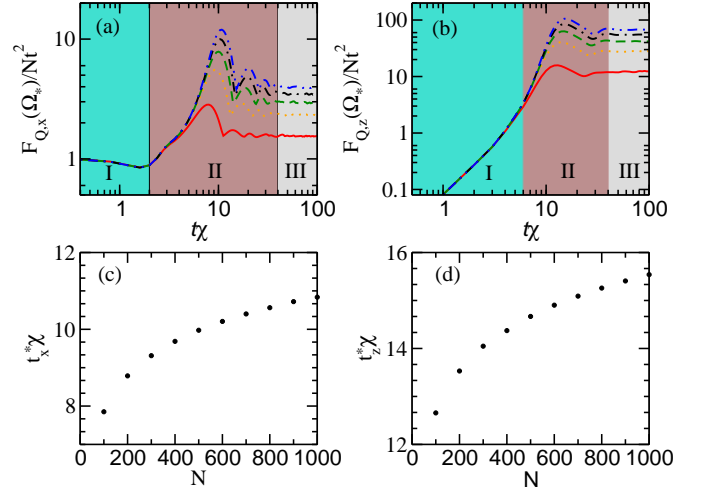}
    \caption{
    The time trace of $F_{Q,x/z}(\Omega^*)/Nt^2$ (a/b) obtained using numerical time propagation 
    and the corresponding critical time $t_{x/z}^{*}$ (c/d) for various system sizes $N$.
    An initial spin coherent state pointing to the south pole on the Bloch sphere and $\omega/\chi=10^{-4}$ are used for the time evolution. In (a/b) the red solid, orange dotted, green dashed, black dot-dashed, and blue double-dot-dashed lines correspond to $N=100, 300, 500, 700,$ and $900$, respectively. 
    The colored regions marked by I, II, and III in (a/b) approximately distinguish the short-, medium-, and long-time dynamical regimes.
    }
    \label{fig:time_trace}
\end{figure}
Insight can also be gained into the short-time behaviour of Eq.~\eqref{F_Q_G}. Using the Taylor expansion 
\begin{eqnarray}
\label{short_time_limit}
\overline{(\hat{\mathcal{H}}_1)}_t \approx \hat{H}_1 + \frac{it}{2} [\hat{H}, \hat{H}_1].
\end{eqnarray}
with $\hat{H}_1=\hat{S}_x$ and $\hat{S}_z$ for perturbations along $\Omega$ and $\omega$, respectively, then the leading order behaviour of the QFI in the short-time limit is most generally
\begin{eqnarray}
\label{short_time_sxz}
\lim_{t\rightarrow 0} F_{Q, x/z}= 4t^2\langle [\Delta \hat{S}_{x/z}(0)]^2\rangle.
\end{eqnarray}
Thus, for an initial state of $\theta =\pi$ we have $F_{Q,x}=Nt^2$ to leading order in time, which is consistent with the results of Fig.~\ref{fig:time_trace}(a) at short times (marked as Region I) where $F_{Q, x}(\Omega^*)/Nt^2 \approx 1$.
In contrast, when considering $F_{Q,x}$ for the same initial state the relevant fluctuations vanish, $\langle [\Delta \hat{S}_{z}(0)]^2\rangle = 0$ and so we must retain higher order contributions in $t$. Specifically, the leading order behaviour of the QFI scales as $t^4$ and is generated by the commutator $it[\hat{S}_z, \hat{H}]/2 = -\Omega\hat{S}_y t/2$ in Eq.~\eqref{short_time_limit}. Then, we obtain $F_{Q,z} = \Omega^2N t^4/4$, which is again in good agreement with the short-time results of Fig.~\ref{fig:time_trace}(b) (see Region I). In particular, $F_{Q,z}(\Omega^*)/Nt^2$ collapses for the range of $N$ considered and shows a power law dependence on $t$.

An intermediate regime separating the long- and short-time limits (labelled as Region II in Fig.~\ref{fig:time_trace}) also exists, wherein $F_{Q, x/z}(\Omega^*)/Nt^2$ shows transient oscillations that depend intimately on the structure of the spectrum and the initial state. Moreover, we identify a critical time $t_{x/z}^*$ where the QFI $F_{Q, x/z}(\Omega^*)/Nt^2$ tends to a maximum value, corresponding to the first peak in Region II.
Figure~\ref{fig:time_trace}(c) and (d) show this critical time $t^*_{x/z}$ as a function of the system size $N$. 
We observe that $t^*_{x/z}$ exhibits only a relatively weak dependence on $N$ and that $t^{*}_{x,z}\chi \sim \mathcal{O}(10)$ in both cases for $N\in [100,1000]$.

\subsection{Approximate model for scaling with system size}
The scaling of the QFI can be directly traced to the emergence a non-analyticity in the energy spectrum of the LMG model. Here, we present a supporting calculation for this discussion. 

Consider Eq.~\eqref{diagonal_2} in the limit of large $N$ such that one can make a continuum approximation $H_1^{nn} \to H_1(E)$ for $E \equiv E_{nn}$. Then, by recognizing that the QFI according to Eq.~\eqref{diagonal_2} is proportional to the characteristic variance of $H_1(E)$, we argue that for an initial state with well defined mean-energy $E$ and energy fluctuations $\Delta E \ll E$ the QFI can be approximated as
\begin{eqnarray}
\label{error_diagonal_ensemble}
F_{Q} = 4t^2\left|\frac{\partial H_1(E)}{\partial E}\right|^2\Delta E^2 . 
\end{eqnarray}

In the case of the LMG model, the divergence of the QFI arises due to a sharp cusp in $S_x(E)$ or kink in $S_z(E)$ at a critical energy $E_{\mathrm{cr}}=N\Omega/2$ [see Figs.~2(b) and (c) in the main text]. Near the critical energy we find both observables are well described by \cite{Perez_2011}
\begin{eqnarray}
\label{fit_model}
\frac{S_{x/z}(E)}{N} = A_{x,z} + B_{x,z} \left(\frac{E}{N}-\frac{E_{\mathrm{cr}}}{N}\right)^{\gamma_{x,z}}.
\end{eqnarray}
where we have used that the energy $E$ and $S_{x/z}(E)$ are extensive observables and thus can be normalized to remove any dependence of $A_{x,z}$, $B_{x,z}$ and $\gamma_{x,z}$ on system size $N$. Substituting Eq.~\eqref{fit_model} into Eq.~\eqref{error_diagonal_ensemble}, evaluating the derivative at $E=E_{\mathrm{cr}}\pm \Delta E$, and using that $\Delta E \sim \sqrt{N}$ for a coherent spin state, we obtain 
\begin{eqnarray}
F^{\mathrm{sec}}_{Q,x/z}\sim 4t^2\gamma_{x,z} B_{x,z} N^{2-\gamma_{x,z}}.
\end{eqnarray}
We then numerically diagonalize the Hamiltonian for a large system ($N = 500$) and fit $S_{x/z}(E)$ using Eq.~\eqref{fit_model} near $E_{\mathrm{cr}}$ to obtain $\gamma_{x} = 0.495\pm0.005$ and $\gamma_z = 0.25\pm0.02$, respectively, where the uncertainties reflect rms error in our fit. To be concrete, we fit $S_{x/z}(E)$ in the energy window of [$E_{\mathrm{cr}}$, $E_{\mathrm{cr}}+n\Delta E$] and [$E_{\mathrm{cr}}-n\Delta E$, $E_{\mathrm{cr}}$] where $n$ is a constant that we tuned through $n=(1,...,5)$ to confirm the estimated scaling parameters $\gamma_{x,z}$ are stable. Example fits are shown in Fig.~(S3). 
\begin{figure}
    \centering
    \includegraphics[scale=0.52]{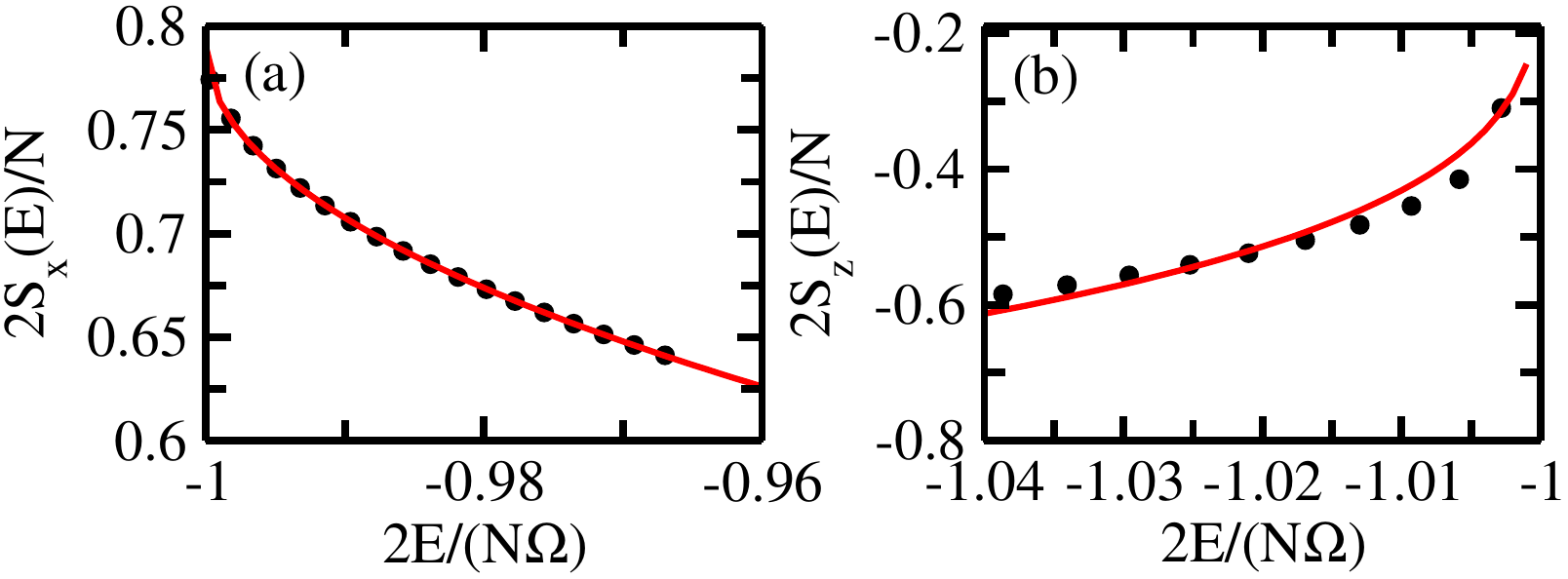}
    \caption{
    The fits of (a) $S_{x}(E)$ and (b) $S_z(E)$ normalized by $N/2$ in the diagonal ensemble near the critical point. The red solid lines correspond to a fit using the formula given in Eq.~\eqref{fit_model}. The relevant fitted parameters are $\gamma_x=0.495\pm0.005$ and $\gamma_z = 0.25\pm0.02$, respectively. Calculations are for $N=500$, $\Omega/\chi=0.5$, and $\omega/\chi=10^{-4}$, but we also check for robustness with $N$. Points within one standard deviation of the total energy for an initial state with $\theta=\pi$ and the same Hamiltonian parameters are used to perform the fitting (see text). 
    }
    \label{fig:sup6}
\end{figure}

As commented in the main text, for the case of the LMG model the sharp features in $S_{x,z}(E)$ are related to a known excited-state quantum phase transition (although these are typically fundamentally distinct phenomena \cite{RLS_Dicke_2021}). Thus, we highlight that in fact we expect the divergence of $S_z(E)$ near $E_{\mathrm{cr}}$ is logarithmic in the thermodynamic limit, identical to the order parameter $\bar{S}_z$. However, our numerical calculations are limited to system sizes where finite size contributions will dominate and it is not possible to distinguish signatures of the logarithmic divergence.

\subsection{Bounds on the QFI}
The standard quantum limit for sensing can be recast in terms of the QFI as $F_Q^{\mathrm{cl}} \leq Nt^2$. This bound is conventionally obtained by considering only quasiclassical initial states that feature no quantum correlations or entanglement and are then subject to evolution under $\emph{only}$ the driving term of $\hat{H}_{\mathrm{LMG}}$, e.g., $\Omega\hat{S}_x$ or $\omega\hat{S}_z$ \cite{Giovannetti_2006,Skotiniotis_2015}. Restricting to spin-$1/2$ systems, coherent spin states satisfy the former condition and a straightforward calculation demonstrates that for suitable choice of the tipping ($\theta$) and azimuthal ($\phi$) angles they saturate $F_Q^{\mathrm{cl}} \leq Nt^2$ where $N$ is the number of spin-$1/2$ particles. 

In the main text we demonstrate that the interplay of interactions $\propto \hat{S}_z^2$ with the driving terms enables one to surpass the SQL near a DPT. Here, we emphasize that this result is a consequence of strong correlations and entanglement in the dynamically generated quantum states, rather than the nonlinearity of the dynamics at the classical level, by explicitly proving that the QFI remains bounded $F_Q \leq Nt^2$ in the mean-field limit. 

Our proof is based upon the form of the QFI given in Eq.~\eqref{F_Q_G}, $F_Q(\lambda,t) = 4t^2\langle [\Delta \overline{(\hat{\mathcal{H}}_1)}_t ]^2 \rangle$. For evolution under any generic Hamiltonian (e.g., $\hat{H} = \hat{H}_0 + \lambda \hat{H}_1$) we can expand $\hat{H}_1(t)$ in terms of products of single-body operators $\hat{h}^{\alpha}_j$,
\begin{eqnarray}
\label{single_particle_exp}
\hat{\mathcal{H}}_1(t) = \sum_{i,\alpha} a^{\alpha}_i(t)\hat{h}^{\alpha}_i + \sum_{i,j,\alpha,\beta} b^{\alpha\beta}_{ij}(t)\hat{h}^{\alpha}_i\hat{h}^{\beta}_j+\cdots ,
\end{eqnarray}
where the index $j$ runs over all particles (sites) and $\alpha$ the set of single-body operators for each particle. For example, $\hat{h}^{\alpha}_j = \hat{\sigma}^{\alpha}_j/2$ with $\alpha \in (x,y,z)$ for an ensemble of $N$ spin-$1/2$ particles indexed by $j=1,...,N$.

Plugging Eq.~\eqref{single_particle_exp} into the definition of $F_Q(\lambda,t)$ [
Eq.~\eqref{F_Q_G}]
we obtain
\begin{align}
    \label{F_corr}
    F_Q=&4t^2\Big[\sum_{i,\alpha} (a^{\alpha}_i)^2\mathrm{Var}(\hat{h}^{\alpha}_i)+\sum_{i\ne j,\alpha,\beta}a^{\alpha}_{i,\alpha}a^{\alpha}_{j,\beta}\mathrm{Cov}(\hat{h}^{\alpha}_i,\hat{h}^{\beta}_j)\\\nonumber
    &+\sum_{i, j, k, \alpha,\beta,\gamma}a_{i}^{\alpha}b_{jk}^{\beta\gamma}\mathrm{Cov}(\hat{h}^{\alpha}_i, \hat{h}_j^{\beta}\hat{h}_k^{\gamma})+\cdots
    \Big],
\end{align}
where $\mathrm{Var}(\hat{O}) = \langle \hat{O}^2 \rangle - \langle \hat{O} \rangle^2$ and $\mathrm{Cov}(\hat{O}, \hat{O}^{\prime}) = \langle \hat{O} \hat{O}^{\prime} \rangle - \langle\hat{O}\rangle\langle\hat{O}^{\prime}\rangle$ correspond to the variance and covariance, respectively. In fact, Eq.~\eqref{F_corr} is equivalent to expanding the QFI with respect to system size $N$ since the first, the second, and the third term inside the square bracket typically scales as $N$, $N^2$, and $N^3$, respectively, which is related to the nature of effective one-, two-, and three-body interactions. 

For an uncorrelated initial state, e.g., $\vert \psi_o \rangle \equiv \vert \Psi_{\mathrm{sp}} \rangle^{\otimes N}$ where $\vert \Psi_{\mathrm{sp}} \rangle$ is some single-particle state, the second term $\sum_{i\ne j}a_i^{\alpha} a_j^{\beta}\mathrm{Cov}(\hat{h}_i^{\alpha}, \hat{h}_j^{\beta})$ vanishes. However, the third term $\mathrm{Cov}(\hat{h}_i^{\alpha}, \hat{h}_j^{\beta}\hat{h}_k^{\gamma})$ can still have non-zero contributions (e.g., for $i=j$, $i=k$, or $j=k$). Nevertheless, if we combine an uncorrelated initial state with the assumption that the Hamiltonian is single-body, i.e., can be decomposed as $\hat{H} \equiv \sum_{j,\alpha} H_{j}^{\alpha} \hat{h}^{\alpha}_j$ then only the linear terms in Eq.~\eqref{single_particle_exp} survive (e.g., $b_{jk}^{\beta\gamma}(t) = 0$ strictly). 

Consequently, $F_Q = 4t^2 \sum_{i,\alpha} (a_i^{\alpha})^2\mathrm{Var}(\hat{h}_i^{\alpha})$.
In general, 
the coefficients and the single-particle variance are bounded by $\sum_{\alpha}(a_i^{\alpha})^2\le 1 $ and $\mathrm{Var}(\hat{h}_i^{\alpha})\le \Delta h^2_{\mathrm{max}}/4$ where $\Delta h_{\mathrm{max}}$ is the difference between the largest and smallest eigenvalues of $\hat{h}$. This leads to $F_Q\le Nt^2\Delta h_{\mathrm{max}}$ \cite{Skotiniotis_2015}. For a spin-1/2 system, we have $\hat{h}_i^{\alpha}=\sigma_i^{\alpha}/2$ and thus $\Delta h_{\mathrm{max}}=1$, leading to the result $F_Q\le Nt^2$.

This discussion is illustrated by using the specific example of the LMG model. We assume an initial (uncorrelated) coherent spin state of $N$ spin-$1/2$ particles and consider the dynamics generated by the mean-field Hamiltonian
\begin{eqnarray}
\label{H_mean_field}
\hat{H}_{\text{MF}} = a(t) \hat{S}_x +b(t) \hat{S}_y +c(t) \hat{S}_z,
\end{eqnarray}
with (time-dependent) coefficients $(a, b, c)=[-\Omega, 0, -2\langle S_z(t)\rangle/N-\omega]$. Rigorously, $\hat{H}_{\text{MF}}$ is the effective Hamiltonian consistent with the equations of motion obtained from $\hat{H}_{\mathrm{LMG}}$ and invoking a mean-field approximation. 

We then compute the QFI for a generic single-body perturbation, $F_{Q,\alpha} = 4t^2\langle [\Delta \overline{(\hat{S}_{\alpha})}_t ]^2 \rangle$. Formally,
\begin{align}
\label{S_x_mean_field}
\hat{S}_{\alpha}(t) =\mathcal{T}\exp\left(i\int_0^t \hat{H}_{\text{MF}}(t') dt'\right) \hat{S}_{\alpha} \nonumber\\
\times \mathcal{T}\exp\left(-i\int_0^t \hat{H}_{\text{MF}}(t') dt'\right),
\end{align}
where $\mathcal{T}$ is the usual time-ordering operator. However, the form of $\hat{H}_{\text{MF}}$ means that Eq.~\eqref{S_x_mean_field} can always be expressed as a simple sum over collective spin operators
\begin{eqnarray}
\label{S_x_t_expand}
\hat{S}_{\alpha}(t) = A_{\alpha}(t) \hat{S}_x + B_{\alpha}(t) \hat{S}_y + C_{\alpha}(t) \hat{S}_z,
\end{eqnarray}
where $A_{\alpha}(t)$, $B_{\alpha}(t)$, and $C_{\alpha}(t)$ are real numbers that satisfy
\begin{eqnarray}
\label{constaint}
A_{\alpha}^2(t) +B_{\alpha}^2(t)+C_{\alpha}^2(t)=1.
\end{eqnarray}

Using Eq.~\eqref{S_x_t_expand} we can thus express the time-average $\overline{(\hat{S}_{\alpha})}_t$ as
\begin{align}
    \label{time_integral_S}
    \int_0^tdt'\hat{S}_{\alpha}(t') = \mathcal{N}_{\alpha}({\bf{n_{\alpha}}}\cdot{\hat{\bf{S}}})
\end{align}
where the normalization factor $\mathcal{N}_{\alpha}$ is
\begin{align}
    \label{normalization}
    \mathcal{N}_{\alpha}^2&=\Bigg[\left(\int_0^tdt'A_{\alpha}(t')\right)^2 +\left(\int_0^tdt'B_{\alpha}(t')\right)^2 \\\nonumber
    &+\left(\int_0^tdt'C_{\alpha}(t')\right)^2\Bigg]\\\nonumber
    & \le\left(\int_0^t dt'\right)\left(\int_0^t\left[A^2_{\alpha}(t')+B^2_{\alpha}(t')+C^2_{\alpha}(t')\right]dt'\right)\\\nonumber
    &=t^2 .
\end{align}
Here, ${\bf{n_{\alpha}}}$ is a unit vector aligned along the axis of the perturbation $\hat{S}_{\alpha}$, and we have used the Cauchy-Scwharz inequality to obtain the second last line. Plugging Eq.~\eqref{time_integral_S} into Eq.~\eqref{F_Q_G} we finally obtain
\begin{align}
\label{Q_F_S_mean_field}
    F_{Q}&=\mathcal{N}_{\alpha}^2\left[\left\langle\left({\bf{n_{\alpha}}}\cdot{\hat{\bf{S}}}\right)^2 \right\rangle - \left\langle{\bf{n_{\alpha}}}\cdot{\hat{\bf{S}}}\right\rangle^2\right]\\\nonumber
    &=\mathcal{N}_{\alpha}^2 \left[1-\left({\bf{n_{\alpha}}}\cdot{\bf{n_{\psi}}}\right)^2\right] N\\\nonumber
    &\le Nt^2
\end{align}
where the average is taken with respect to an arbitrary coherent spin state polarized in the direction of the unit vector ${\bf{n_{\psi}}}$. This result emphasizes that the nonlinear dynamics demonstrated by the classical model are not sufficient to generate the large QFI we observe in the full quantum dynamics, and instead this arises because of complex features that are generated in the quantum noise (see, e.g., Fig.~3(a) of the main text).

\begin{figure*}[tbh]
    \centering
    \includegraphics[width=16cm]{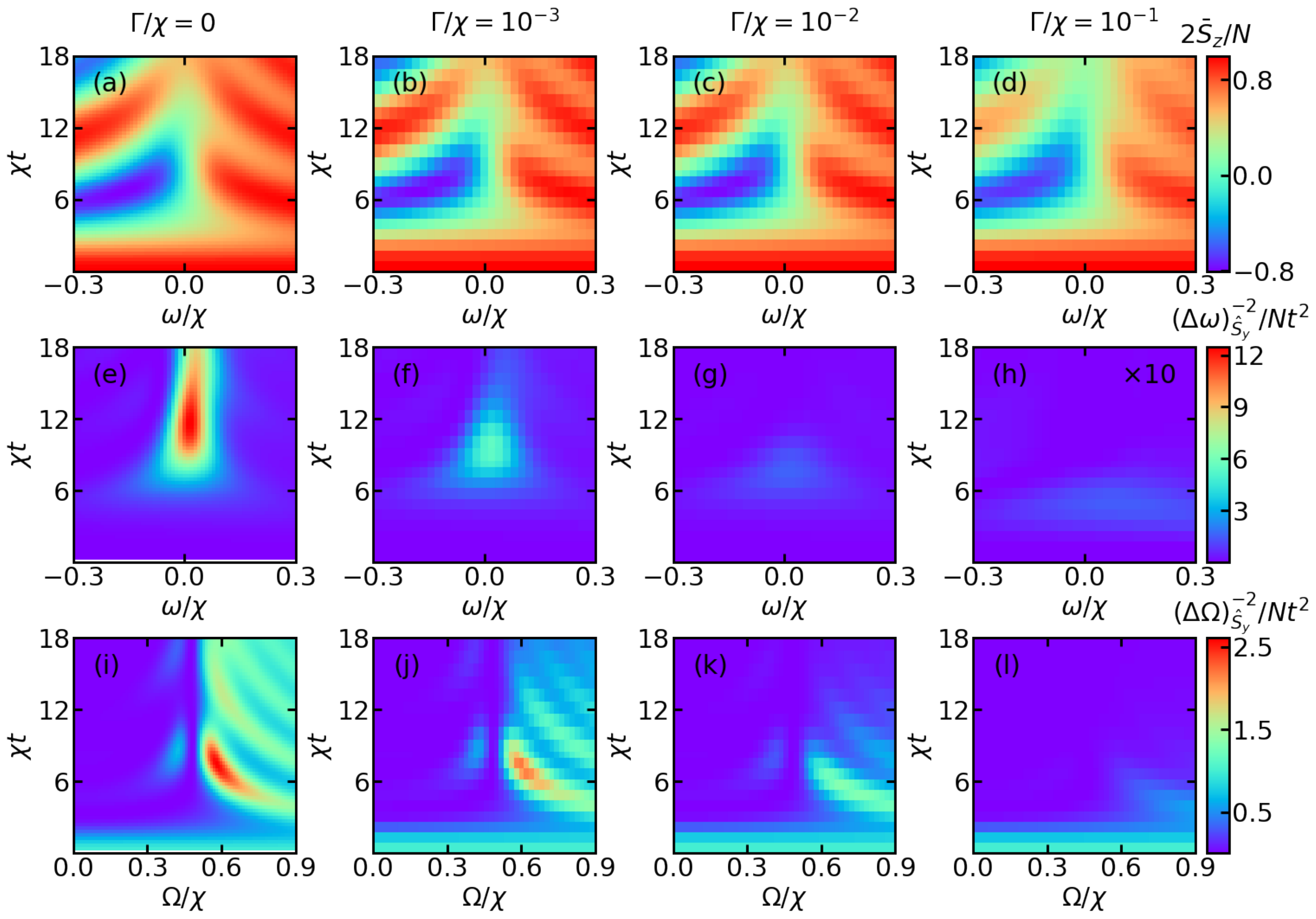}
    \caption{(a) Time-averaged order parameter $\bar{S}_z$ and (b)-(c) scaled inverse sensitivities $(\Delta\omega)_{\hat{S}_y}^{-2}/Nt^2$ and $(\Delta\Omega)_{\hat{S}_y}^{-2}/Nt^2$ as a function of time and longitudinal/transverse field strengths $\omega/\chi$ and $\Omega/\chi$ including decoherence at rate $\Gamma/\chi$ (indicated in panels). Calculations are performed using an efficient exact numerical solution of the master equation (see text) with parameters $N=100$ and $\Omega/\chi = 1/2$.
    Note the $z$-value of Panel (h) is multiplied by a factor of 10 to make it visible.}
    \label{fig:sup4}
\end{figure*}

\section{Numerical methods}

\subsection{Closed system}
We numerically simulate the dynamics of the closed system governed by the Hamiltonian $\hat{H}_{\mathrm{LMG}}$ [Eq.~(3) of the main text] using an efficient Chebyshev scheme. In this method, an arbitrary time-evolved state, $\vert \psi(t) \rangle = \hat{U}(t) \vert \psi_0 \rangle$ where $\hat{U}(t) = e^{-i\hat{H}_{\mathrm{LMG}}t}$, is obtained by expanding the time propagator into a superposition of Chebyshev polynomials $\phi_n$ for a single time step \cite{Kosloff_JCP1984}:  
\begin{align}
\label{chebyshev}
\hat{U}(t) \approx e^{-i(E_{\mathrm{max}}-E_{\mathrm{min}})t/2}\sum_{n=0}^{N_{\mathrm{cut}}} a_n(t)\phi_n(-i\hat{H}_{\mathrm{norm}}) .
\end{align}
Here, we have introduced the normalized Hamiltonian,
\begin{eqnarray}
\label{normalized_H}
\hat{H}_{\mathrm{norm}} = \frac{\hat{H}-(E_{\mathrm{max}}+E_{\mathrm{min}})/2}{E_{\mathrm{max}}-E_{\mathrm{min}}} ,
\end{eqnarray}
and the expansion coefficients $a_n(t)$ are given by 
\begin{eqnarray}
\label{coeff}
a_n(t) = 
\begin{cases}
2 J_n\left(\frac{(E_{\mathrm{max}}-E_{\mathrm{min}})t}{2}\right) \text{for}\ n>0 ,\\
J_0\left(\frac{(E_{\mathrm{max}}-E_{\mathrm{min}})t}{2}\right) \text{for}\ n=0 ,
\end{cases}
\end{eqnarray}
where $J_n$ is the $n$th Bessel function. 
The free parameters $E_{\mathrm{min}}$ and $E_{\mathrm{max}}$ are chosen such that the spectrum of $\hat{H}$ is appropriately encompassed by the energy window $[E_{\mathrm{min}}, E_{\mathrm{max}}]$. Throughout the manuscript we choose $E_{\mathrm{max}} = N(\chi+|\omega|)$ and $E_{\mathrm{min}} = -N(\chi+|\omega|)$.

To efficiently construct the complex Chebyshev polynomial $\phi_n(-i\hat{H}_{\mathrm{norm}})=(-i)^{n}T_n(\hat{H}_{\mathrm{norm}})$
where $T_n(x)$ is the Chebyshev polynomial of the first kind, 
we use the recursion relation
\begin{multline}
\label{recursive}
\phi_{n+1}(-i\hat{H}_{\mathrm{norm}}) = -2i\hat{H}_{\mathrm{norm}}\phi_{n}(-i\hat{H}_{\mathrm{norm}})\\+\phi_{n-1}(-i\hat{H}_{\mathrm{norm}}) ,
\end{multline}
with the initial condition $\phi_0 = 1$ and $\phi_1 = -i\hat{H}_{\mathrm{norm}}$. 

The Chebyshev expansion is expected to converge exponentially with increasing $N_{\mathrm{cut}}$ provided the $N_{\mathrm{cut}}$ is not less than $(E_{\mathrm{max}}-E_{\mathrm{min}})t/2$. To safely satisfy this requirement we also choose $N_{\mathrm{cut}}$ to exceed this theoretical value by $30\%$. We check convergence by computing the normalization of the wavefunction $\vert \langle \psi(t) \vert \psi(t) \rangle \vert^2$ and ensure that it deviates from unity by less than $10^{-10}$
for all runs. Importantly, this deviation is much smaller than any perturbation to the wavefunction introduced by $\delta\Omega$ or $\delta\omega$ when computing the QFI.

\subsection{Open system with decoherence}

The dynamics of the LMG model in the presence of single-particle decoherence can be efficiently simulated by exploiting the permutation symmetry of the model, combined with the fact that the initial states we probe are fully collective (i.e., $\langle \hat{S}^2 \rangle = \frac{N}{2}(\frac{N}{2} + 1)$ for our chosen initial states). In generality, the dynamics of the open system are described by a master equation for the density matrix $\hat{\rho}$ of the spin ensemble \cite{Foss_Feig_2013},
\begin{equation}
    \frac{d\hat{\rho}}{dt} = -i\left[ \hat{H},\hat{\rho} \right] + \frac{\Gamma}{4} \sum_{j=1}^N \left( \hat{\sigma}^z_j\hat{\rho} \hat{\sigma}^z_j - \hat{\rho} \right) \label{eqn:masterEqn}
\end{equation}
where $\hat{H}$ is the LMG Hamiltonian [see Eq.~(3) of the main text]. To efficiently solve Eq.~\eqref{eqn:masterEqn} we exploit the permutation symmetry of both the Hamiltonian and dissipative terms to reduce the scaling of problem from $4^N$ to $\mathcal{O}(N^3)$. This enables us to exactly (up to numerical precision) compute the dissipative dynamics of systems up to $N \sim \mathcal{O}(100)$ with relative ease, enabling meaningful comparisons with current state-of-the-art AMO quantum simulators. A full analysis and discussion of this method can be found in Refs.~\cite{Geremia_2010,Geremia_2008} and citations therein.

In Fig.~\ref{fig:sup4} we compare the behaviour of the time-averaged order parameter $\bar{S}_z$ and scaled inverse sensitivity $(\Delta\omega)^{-2}_{\hat{S}_y}/Nt^2$ across a range of longitudinal field strengths $\omega/\chi$ and decoherence rates $\Gamma/\chi$. We observe that $\bar{S}_z$ is relatively robust to $\Gamma$, as the primary effect of dephasing is to damp out oscillations in the disordered phase, consistent with recent experimental observations \cite{Muniz2020}. The sensitivity is more noticeably degraded by decoherence, particularly beyond short time scales ($\gamma t \ll 1$). Nevertheless, we observe that the transient peak of $(\Delta\omega)^{-2}_{\hat{S}_y}/Nt^2$ is preserved, albeit shifted towards earlier time scales and gradually smeared out around the transition. For large $\Gamma/\chi \sim 0.1$ we observe that the peak becomes systematically shifted away from the ideal DPT at $\omega/\chi = 0$ (consistent with the behaviour of $\bar{S}_z$), but it remains clearly centered near $\omega/\chi = 0$ for weaker decoherence. This last comparison is not unexpected, as the most noticable features of the DPT in the QFI arise for $\chi t \gtrsim 10$ and thus the naive requirement $\gamma t \ll 1$ for decoherence to be perturbative translates to the condition $\gamma/\chi \ll 0.1$ for the QFI to display robust transient signatures.

\begin{figure}[tbh]
    \centering
    \includegraphics[width=9cm]{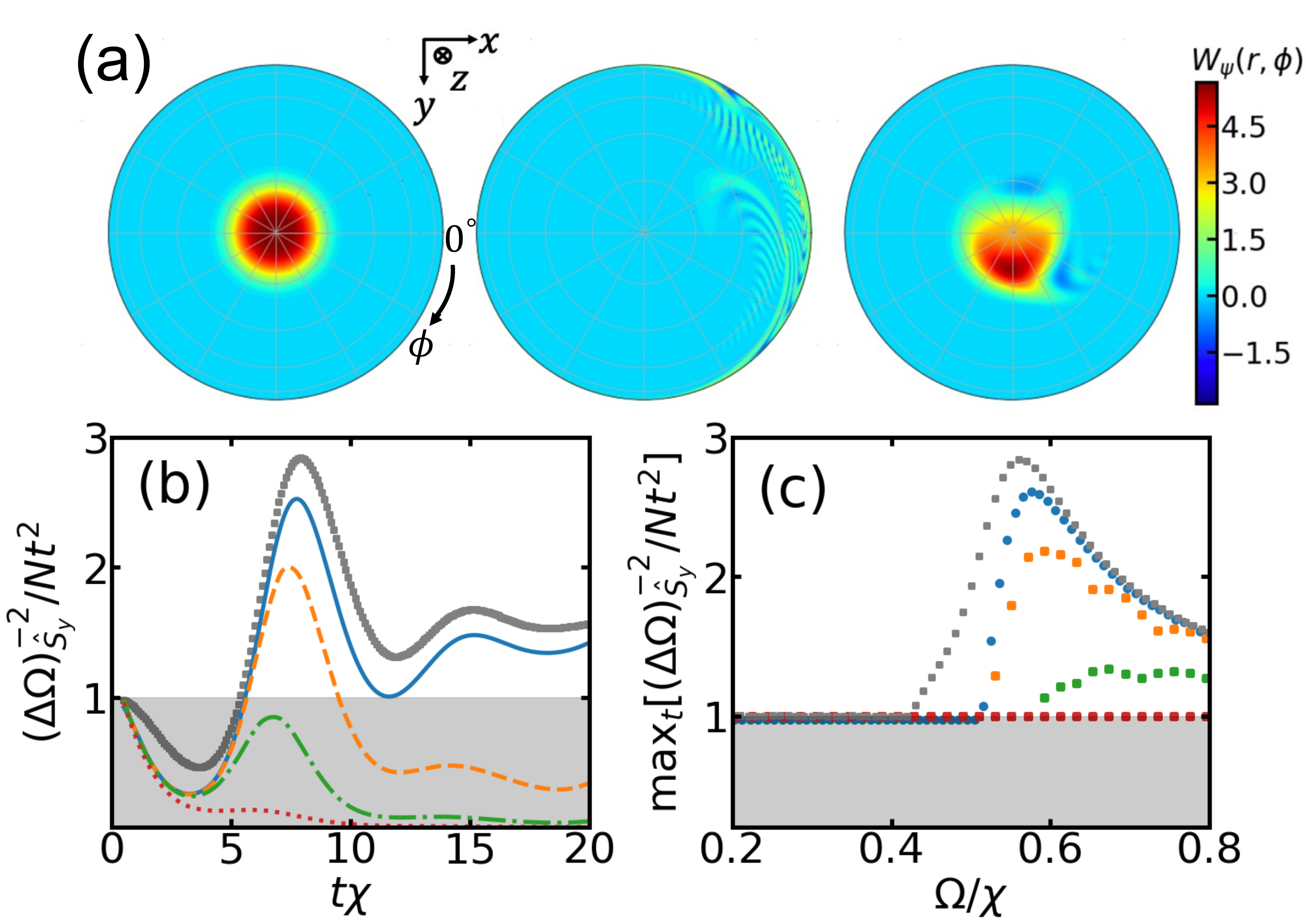}
    \caption{(a) Typical Wigner functions \cite{WignerDis1994} $W_{\psi}(r, \phi)$ of the initial ($\vert \psi_0\rangle$), intermediate ($\vert \psi(t) \rangle$) and final ($\vert \psi_f \rangle$) states for $\chi t = 8$, $\omega = 0$ and $\delta\Omega/\chi = 7\times10^{-3}$. We plot with polar coordinates $r=\left(1+2S_z/N\right)^{1/4}$ and $\phi = \mathrm{atan}(S_y/S_x)$. (b) Normalized inverse sensitivity $(\Delta\Omega)^{-2}_{\hat{S}_y}/(Nt^2)$ for the same parameters as (a) [$\Omega/\chi=0.556$]. The blue solid, orange dashed, green dot-dashed, and red dotted lines correspond to $\Gamma/\chi = 0, 10^{-3}, 10^{-2}$, and $10^{-1}$, respectively. For comparison we also plot the normalize QFI $F_{Q,x}/(Nt^2)$, and indicate the sensitivity regime bounded by the normalized SQL, $(\Delta\omega)^{-2}_{\mathrm{SQL}}/Nt^2 = 1$, by gray shading. (c) Maximum of the normalized inverse sensitivity, $\mathrm{max}_t[(\Delta\omega)^{-2}_{\hat{S}_y}/Nt^2]$, as a function of transverse field $\Omega/\chi$ for a range of decoherence rates $\Gamma/\chi$ [same color coding as (b)]. We again indicate $\mathrm{max}_t[F_{Q,x}/Nt^2]$ and the SQL. All data is for $N=100$ with an initial state of all spins aligned along $-\hat{z}$. 
    }  
    \label{fig:sup5}
\end{figure}

Calculations for a perturbation of the transverse field yields similar results. As shown in Fig.~\ref{fig:sup5} we again observe that the final state after the echo is well distinguished by measurements of $\hat{S}_y$. Panel (a) illustrates the Wigner distribution $W_{\psi}(r,\phi)$~\cite{WignerDis1994}, which shows qualitatively similar features to Fig.~3 of the main text although the final state is instead typically displaced along the $+y$-direction. Similarly, the inverse sensitivity tracks the dynamics of the QFI [panel (b)]. However, we also note that in this case we always find the maximum (transient) sensitivity is at least as good as the SQL [panel (c)], $\mathrm{max}_t[(\Delta\Omega)^{-2}_{\hat{S}_y}/Nt^2] \geq 1$, as for $\chi t \to 0$ the perturbation results merely in a rotation of a simple coherent spin state, which is precisely the operational definition of the SQL. Nevertheless, a pronounced peak of $\mathrm{max}_t[(\Delta\Omega)^{-2}_{\hat{S}_y}/Nt^2]$ still reflects the underlying DPT for weak decoherence, while the dynamical phases are still delineated by $\mathrm{max}_t[(\Delta\Omega)^{-2}_{\hat{S}_y}/Nt^2] = 1$ (ordered) and $\mathrm{max}_t[(\Delta\Omega)^{-2}_{\hat{S}_y}/Nt^2] > 1$ (disordered), respectively. A complete examination of the inverse sensitivity $(\Delta\Omega)^{-2}_{\hat{S}_y}$ for varied $\Omega$ and $\chi t$ is also shown in Fig.~\ref{fig:sup4}.

\subsection{Scaling of the sensitivity with system size}
To verify a robust correspondence between the QFI and the sensitivities obtained via the echo, $(\Delta\Omega)_{\hat{S}_y}$ and $(\Delta\omega)_{\hat{S}_y}$, we compute the scaling of both quantities as a function of system size. At a fixed long time $\chi t = 10^3$ we fit the computed inverse sensitivity to the function $aN^b$ and obtain $(\Delta\Omega)^{-2}_{\hat{S}_y} \sim N^{1.4}$ and $(\Delta\omega)^{-2}_{\hat{S}_y} \sim N^{1.78}$ across a window of $N\in [100, 2000]$. These results closely follow the scaling of the QFI obtained via integrating the full quantum dynamics, $F_{Q,x} \sim N^{1.5}$ and $F_{Q,z}\sim N^{1.78}$, extracted with the same procedure. 
We note that $(\Delta\omega)^{-2}_{\hat{S}_y}$ has the same system size scaling compared to $F_{Q,z}$
but with a smaller prefactor.
As for $(\Delta\Omega)^{-2}_{\hat{S}_y}$,
both the prefactor and the system size scaling are less than those for $F_{Q,x}$.

\section{Classical dynamical phase diagram}
The dynamical phase diagram of the LMG model can be solved analytically in the classical ($N\to \infty$) limit (see, e.g., Refs.~\cite{lerose2019dpt,Muniz2020,RLS_Dicke_2021}). Briefly, the classical limit is equivalent to solving the equations of motion for expectation values $\langle \hat{S}_{x,y,z} \rangle$ under a mean-field approximation wherein all higher-order correlations are expressed as the product of single-body terms, e.g., $\langle \hat{O}_1 \hat{O}_2 \rangle = \langle \hat{O}_1 \rangle \langle \hat{O}_1 \rangle$. Assuming an initial state where all the spins are fully polarized along an arbitrary axis, i.e., $\mathbf{S} \equiv \left( \langle \hat{S}_x \rangle, \langle \hat{S}_y \rangle, \langle \hat{S}_z \rangle \right) = \left( \frac{N}{2}\mathrm{sin}(\theta)\mathrm{cos}(\phi), \frac{N}{2}\mathrm{sin}(\theta)\mathrm{sin}(\phi), \frac{N}{2}\mathrm{cos}(\theta) \right)$, the dynamics of the mean-field observable $S_z(t)=\langle\hat{S_z}(t)\rangle$ can be reduced to an equivalent model of a classical particle in a potential, described by the differential equation:
\begin{eqnarray}
\label{eom}
\frac{1}{2}\left(\frac{dS_z(t)}{dt}\right)^2 + V_{\mathrm{eff}}(S_z) = 0 .
\end{eqnarray}
Here, the effective potential $V_{\mathrm{eff}}$ is
\begin{align}
 \label{eff_pot}
 V_{\mathrm{eff}}(S_z)=\frac{1}{2}\left(E+\frac{\chi}{N}S_z^2+\omega S_z\right)^2+\frac{\Omega^2S_z^2}{2}-\frac{\Omega^2N^2}{8} ,
\end{align}
and the total energy
\begin{align}
\label{initial_energy}
E = -\frac{N}{2}\left[\frac{\chi}{2}\cos^2(\theta) + \Omega\sin(\theta)\cos(\phi)+\omega\cos(\theta)\right],
\end{align}
is a conserved quantity. 

\begin{figure}[tbh]
    \centering
    \vspace{0.2cm}
    \hspace{-0.5cm}
    \includegraphics[scale=0.31]{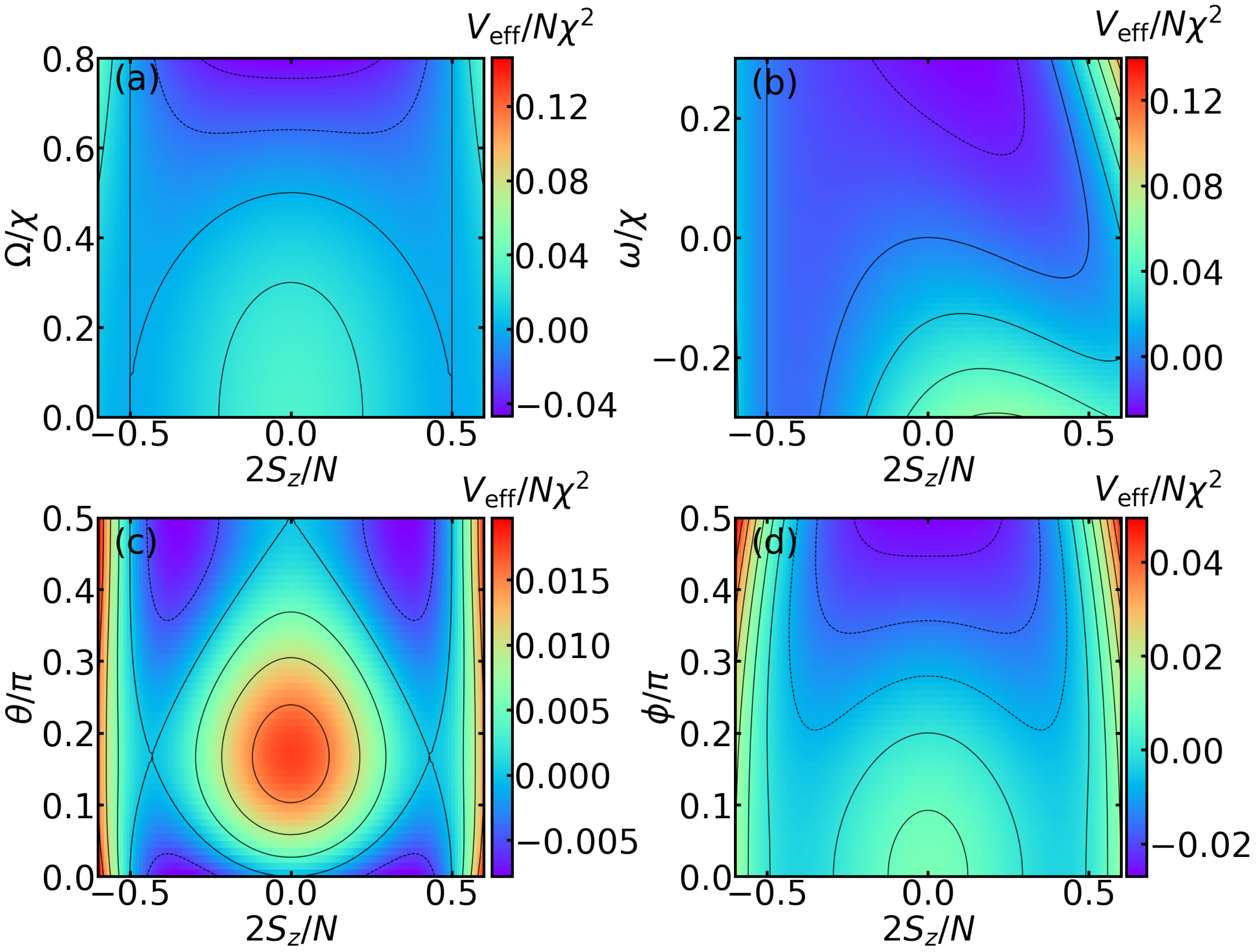}
    \caption{
    The 2D plot of the effective potential $V_{\mathrm{eff}}(S_z)/\chi^2N^2$ as functions of (a) $\Omega$ and $S_z$ for $\theta=\pi$, $\phi=0$, and $\omega=0$, (b) $\omega$ and $S_z$ for $\theta=\pi$, $\phi=0$, and $\Omega=0.5\chi$, (c) $\theta$ and $S_z$ for $\phi=0$, $\Omega/\chi=0.5$, and $\omega=0$, and (d)
     $\phi$ and $S_z$ for $\theta=0.3\pi$, $\Omega/\chi=0.5$, and $\omega=0$.}
    \label{fig:sup1}
\end{figure}

Figure~\ref{fig:sup1} shows various cuts of $V_{\mathrm{eff}}(S_z)/\chi^2N^2$ for selected parameter combinations of $\theta, \phi, \Omega/\chi$, and $\omega/\chi$.  For all the cases shown in Fig.~\ref{fig:sup1}, a transition from a double well to a single well can be seen. In the double-well regime, a local potential barrier exists between the two wells, which supports a local maximum at $S_z^*$. The relation of the total mechanical energy of the initial state in comparison to the magnitude of the potential barrier controls the dynamical phase: The ordered phase corresponds to the case where the particle is confined to a single well, whereas the disordered phase corresponds to the case where the particle has sufficient energy to traverse the barrier and oscillate between both wells. For an initial state with $S_z \neq 0$ the transition between different dynamical phases can be obtained as the condition for which the classical turning point of the particle coincides with $S_z^*$, i.e., $V_{\mathrm{eff}}(S_z^*) = 0$. 

In general, obtaining an analytic solution for the phase boundary is non-trivial, due to the quartic nature of $V_{\mathrm{eff}}(S_z)$. However, analytical expressions can be obtained in two special cases. First, for $\omega=0$ the local maximum occurs at $S_z^*=0$ due to a parity symmetry of the model (i.e., the dynamics is unchanged upon the transformation $S_z \rightarrow -S_z$ and $S_x\rightarrow S_x$), enabling a straightforward solution of the critical transverse field \cite{RLS_Dicke_2021},
\begin{eqnarray}
\label{omega_cr_delta0}
\Omega_{\mathrm{cr}} =  \pm\frac{0.5\cos^2(\theta)}{1\mp \sin(\theta)\cos(\phi)}.
\end{eqnarray}
Second, for $\theta=\pi$ or $\theta=0$ the dependence on the azimuthal angle $\phi$ is eliminated and we obtain an expression for the critical transverse field as a function of the longitudinal field \cite{Muniz2020},
\begin{align}
\label{omega_cr_theta_pi}
\Omega_{\mathrm{cr}} &= \frac{\chi}{2}\Bigg[2\left(1-\frac{\omega}{\chi}\right)\left(1+\frac{2\omega}{\chi}\right)\\\nonumber
&-\frac{3}{2}\left(\frac{8\omega}{\chi}+1\right)+\frac{1}{2}\left(1+\frac{8\omega}{\chi}\right)^{3/2}\Bigg]^{1/2}.
\end{align}

\bibliography{library}

\end{document}